\documentclass[usenatbib]{mn2e}
\usepackage{epsfig}
\usepackage{amssymb}
\usepackage{lscape,graphicx}
\usepackage{longtable}
\usepackage{psfig}
\usepackage{lscape}
\usepackage{rotating}
\usepackage{graphicx}
\usepackage{epstopdf}

\def\gsim{\mathrel{\raise0.35ex\hbox{$\scriptstyle >$}\kern-0.6em
\lower0.40ex\hbox{{$\scriptstyle \sim$}}}}
\def\lsim{\mathrel{\raise0.35ex\hbox{$\scriptstyle <$}\kern-0.6em
\lower0.40ex\hbox{{$\scriptstyle \sim$}}}}

\begin{document}
\title[Gas and BHs of QSOs and SMGs at $z\sim2$]{Testing the evolutionary link between submillimetre galaxies and quasars: CO observations of QSOs at $z\sim2$}

\author[Coppin et al.]{ 
\parbox[t]{\textwidth}{
K.\,E.\,K.\ Coppin$^{1}$, A.\,M.\ Swinbank$^{1}$, R.\ Neri$^{2}$, P.\ Cox$^{2}$,  D.\,M.\ Alexander$^{3}$, Ian Smail$^{1}$, M.\,J.\ Page$^{4}$, J.\,A.\ Stevens$^{5}$, K.\,K.\ Knudsen$^{6}$, R.\,J.\ Ivison$^{7,8}$, A.\ Beelen$^{9}$, F.\ Bertoldi$^{6}$, A.\ Omont$^{10}$}\\\\
$^{1}$ Institute for Computational Cosmology, Durham University, South Road, Durham, DH1 3LE\\
$^{2}$ Institut de RadioAstronomie Millim\'{e}trique (IRAM), 300 rue de la Piscine, Domaine Universitaire, 38406 Saint Martin d'H\`{e}res, France\\
$^{3}$ Department of Physics, Durham University, South Road, Durham, DH1 3LE\\
$^{4}$ UCL, Mullard Space Science Laboratory, Holmbury St. Mary, Dorking, Surrey RH5 6NT\\
$^{5}$ Centre for Astrophysics Research, University of Hertfordshire, College Lane, Hatfield, AL10 9AB\\
$^{6}$ Argelander-Institut f\"{u}r Astronomie, University of Bonn, Auf dem H\"{u}gel 71, D-53121 Bonn, Germany\\
$^{7}$ Institute for Astronomy, University of Edinburgh, Royal Observatory, Blackford Hill, Edinburgh, EH9 3HJ\\
$^{8}$ UK Astronomy Technology Centre, Royal Observatory, Blackford Hill, Edinburgh, EH9 3HJ\\
$^{9}$ Institut d'Astrophysique Spatiale, Universit\'{e} Paris-Sud, F-91405 Orsay, France\\
$^{10}$ Institut d'Astrophysique de Paris, Universit\'{e} Pierre \& Marie Curie, 98 bis Boulevard Arago, F-75014 Paris, France\\
}
\maketitle
\begin{abstract}
We have used the IRAM Plateau de Bure 
millimeter interferometer and the UKIRT 1--5\,$\mu$m Imager Spectrometer (UIST) 
to test the connection between the major phases of 
spheroid growth and nuclear accretion by mapping CO emission in nine submillimetre-detected QSOs at $z=1.7$--2.6 
with black hole (BH) masses derived from near-infrared spectroscopy.
When combined with one QSO obtained from the literature, we present sensitive CO(3--2) or CO(2--1) 
observations of 10 submillimetre-detected QSOs selected at the epoch of peak activity 
in both QSOs and submillimetre (submm) galaxies (SMGs).  
CO is detected in 5/6 very optically luminous ($M_\mathrm{B}\sim -28$)
submm-detected QSOs with BH masses
$M_\mathrm{BH}\simeq10^9$--10$^{10}$\,M$_\odot$, confirming the
presence of large gas reservoirs of $M_\mathrm{gas}\simeq3.4\times10^{10}$\,M$_\odot$.   Our BH masses and dynamical mass constraints on
the host spheroids suggest, at face value, that 
 these optically luminous QSOs at $z=2$ lie about an order of magnitude above the local BH-spheroid
 relation, $M_\mathrm{BH}/M_\mathrm{sph}$, although this result is dependent on the size and inclination of the CO-emitting region.  
However, we find that their BH masses are $\sim30$ times too large and their surface density is
 $\sim300$ times too small to be related to typical SMGs in an
 evolutionary sequence.  Conversely, we measure weaker CO emission in
 four fainter ($M_\mathrm{B}\sim-25$) submm-detected QSOs with properties, BH masses ($M_\mathrm{BH}\simeq5\times10^8$\,M$_\odot$), and
 surface densities similar to SMGs.  These QSOs appear to lie near the local
 $M_\mathrm{BH}/M_\mathrm{sph}$ relation, making them plausible
 `transition objects' in the proposed evolutionary sequence linking
 QSOs to the formation of massive young galaxies and BHs at high-redshift.  We show that SMGs have a
higher incidence of bimodal CO line profiles than seen in our QSO
sample, which we interpret as an effect of their relative inclinations, with the QSOs seen more face-on. 
Finally, we find that the gas masses of the four fainter submm-detected QSOs 
imply that their star formation episodes could be sustained for $\sim10$\,Myr, 
and are consistent with representing a phase in the formation of massive galaxies which
overlaps a preceding SMG starburst phase, 
before subsequently evolving into a population of present-day massive ellipticals.  
\end{abstract}


\section{Introduction}

It has been established that every massive, local
spheroid harbours a supermassive black hole (SMBH) in its 
centre whose mass is proportional to that
of its host (e.g.\ \citealt{Magorrian98}; \citealt{Gebhardt00}). This
suggests that the black holes (BHs) and their surrounding galaxies
were formed synchronously.  This hypothesis has found support from
hydrodynamical simulations of galaxy formation, which use feedback from
winds and outflows from active galactic nuclei (AGN) to link the growth
of the SMBH to that of its host (e.g.\ \citealt{DiMatteo05}; 
\citealt{Hopkins05}; \citealt{Bower06}). 
Thus these models support a picture, first presented by
\citet{Sanders88}, where a starburst-dominated ultra-luminous
infrared galaxy (ULIRG), arising from a merger, evolves first into an
obscured QSO and then into an unobscured QSO, before finally becoming a
passive spheroid.

The high-redshift population of ULIRGs in this proposed evolutionary cycle are the 
submillimetre (submm) galaxies (SMGs; \citealt{SIB97}; \citealt{Chapman05}; 
\citealt{Coppin06}).  These systems have
ULIRG-like bolometric luminosities, $L_\mathrm{IR}\geq 10^{12}$\,L$_\odot$ 
(\citealt{Kovacs06}; \citealt{Coppin07}), and
they have many of the properties expected for gas-rich mergers
(\citealt{Swinbank04,Swinbank06}; \citealt{Tacconi06}).  This population
evolves rapidly out to a peak at $z\sim 2.3$, crudely matching
the evolution of QSOs \citep{Chapman05} and providing
additional circumstantial evidence for a link between SMBH growth and
spheroids.  Two further results have shed light on the evolutionary
link between SMGs and QSOs.  Firstly, a modest fraction of optically luminous QSOs at $z\sim 2$ are
detected in the submm/mm ($\sim(25\pm10)$\%; \citealt{Omont03}) showing that
the QSO- and SMG-phases do not overlap significantly, given the lifetime estimates of the two populations 
(QSOs make up $\sim 4$\% of flux-limited samples of SMGs; \citealt{Chapman05}).  But
when a QSO {\it is} detected in the submm/mm then it could to be in
the transition phase from an SMG to an unobscured QSO, making its
properties a powerful probe of the evolutionary cycle 
(e.g.\ \citealt{Stevens05}; \citealt{Page04}).  Secondly, 
the evolutionary state of the SMBHs within SMGs can also be judged using 
the 2-Ms {\it Chandra} Deep Field North 
observations \citep{Alexander03} to derive
accurate AGN luminosities and hence lower limits on the BH masses ($M_\mathrm{BH}$) in
those SMGs with precise redshifts in this region 
(\citealt{Alexander05nat,Alexander05}; 
\citealt{Borys05}).  These studies suggest that the AGN in
typical SMGs are growing almost continuously -- but that the SMBHs in
these galaxies appear to be several times less massive
than seen in comparably massive galaxies at $z\sim 0$ \citep{Alexander07}.

Together these results argue for a fast transition from a 
star-formation-dominated SMG-phase to the AGN-dominated
QSO-phase \citep{Page04}. The latter phase will result in the 
rapid BH growth necessary to
account for the present-day relation between spheroid and SMBH masses 
(e.g.\ \citealt{Magorrian98}; \citealt{Gebhardt00}).
{\it Can we confirm this and more generally test the proposed
evolutionary link between SMGs and QSOs at the peak of their activity
at $z\sim 2$?}

This evolutionary cycle has been tested in the local Universe by comparing the properties of
QSOs and ULIRGs (e.g.\ \citealt{Tacconi02}).
However, both of these populations are $1000\times$ less abundant in the 
local Universe than they were at the era of their peak activity at $z\sim2$ and so we have to be cautious
about extrapolating from local examples to the high-redshift
progenitors of the bulk of today's massive spheroids (\citealt{Genzel03}; 
\citealt{Swinbank06}). Thus, to properly test the validity of
this cycle for typical spheroids we have to compare QSOs and ULIRGs at
the era where their populations peaked:  $z\sim2$.

The critical pieces of information needed
to test the link between SMGs and QSOs are the relative dynamical,
gas and SMBH masses of these two populations.  
In principle the dynamical masses can be derived from optical
or near-infrared observations of emission line gas in the SMGs or QSOs 
(see \citealt{Swinbank04,Swinbank05,Swinbank06}).
However, there is the problem of removing the QSO
emission in these observations, as well as the effects of dust obscuration and outflows.  
In contrast, molecular CO emission line profiles
are relatively immune to the effects of obscuration and outflows, while
at the same time yielding additional constraints on the relationship
between QSOs and SMGs from their gas masses.

There is currently a lack of sensitive CO observations of QSOs at
$z\sim 1$--3, with data published on only eight sources (e.g.~\citealt{Frayer98}; 
\citealt{Guilloteau99}; \citealt{Beelen04}; \citealt{Hainline04}).  
Instead the focus has been on CO studies of QSOs at $z\gsim4$ 
(e.g.\ \citealt{Omont96}; \citealt{Walter04}; \citealt{Riechers06}), although these 
QSOs have little overlap with the redshift range where SMGs are typically detected.  
The paucity of CO constraints for $z>1$ QSOs also reflects
the difficulty in determining their systemic redshifts with sufficient
precision to guarantee that the CO emission falls within the
bandwidth of typical millimetre (mm) correlators.  However, sensitive
near-infrared spectroscopy of the C{\sc iv}, Mg{\sc ii}, and [O{\sc iii}]5007 emission lines in
QSOs can provide redshifts with required precision as well as H$\alpha$ or H$\beta$ fluxes and 
line widths to yield $M_\mathrm{BH}$ estimates (\citealt{Takata07}; \citealt{Alexander07}).

We have carried out a quantitative test of the proposed link between
SMGs and QSOs at $z\sim 2$ where both populations are most common.  We
have obtained precise systemic redshifts from near-infrared spectroscopy 
of potential transition QSOs (i.e.~submm/mm-detected QSOs) and 
then used the IRAM Plateau de Bure Interferometer (PdBI) to search for 
CO emission.  We relate their dynamical, gas and SMBH masses to SMGs
from the PdBI CO survey \citep{Greve05}.  
We test: a) whether the cold gas
masses in these QSOs are similar to those in SMGs; b) whether the line widths and dynamical
masses of these two populations are comparable; and c) how the
ratio of SMBH to dynamical masses for these submm-detected QSOs relate to 
the estimates for SMGs and those for optically luminous QSOs (which lie
on the present-day $M_\mathrm{BH}$--$M_\sigma$ relation; \citealt{McLure04}).  
Together these observations can 
constrain the proposed evolutionary sequence which links QSOs to the
formation of massive young galaxies and SMBHs at high redshift.

We describe the sample selection, observations and data reduction in \S \ref{obsdr}.  
The results of the near-infrared and mm CO spectra are given in \S \ref{results}.
The CO properties of the submm-detected QSOs are compared and contrasted with SMGs in \S \ref{discuss}, and we discuss the 
evolutionary status of the submm-detected QSOs in \S \ref{evolution}.  Our conclusions are given in \S \ref{concl}.
We adopt cosmological parameters from the \textit{WMAP} fits in \citet{Spergel03}: 
$\Omega_\Lambda=0.73$, $\Omega_\mathrm{m}=0.27$, and
$H_\mathrm{0}=71$\,km\,s$^{-1}$\,Mpc$^{-1}$.  All quoted magnitudes are 
on the Vega system.

\setcounter{figure}{0}

\begin{figure}
\epsfig{file=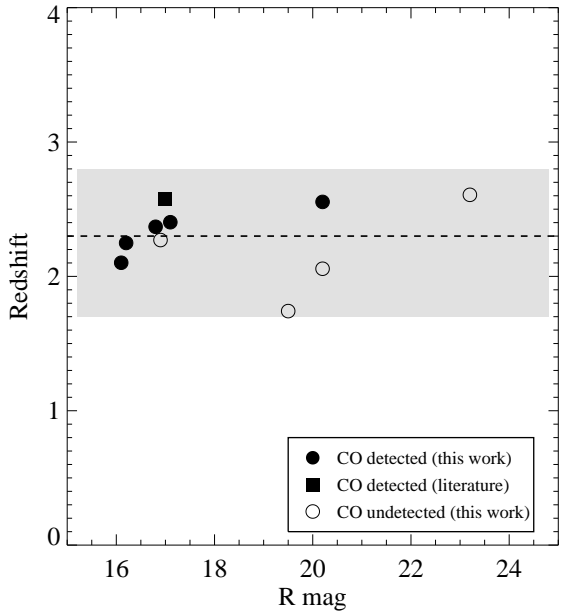,width=0.5\textwidth}
\caption{$R$-band apparent magnitude versus redshift for our sample of submm-detected QSOs.  We show that these lie within the interquartile range (shaded region) of the redshift distribution of SMGs (median $<\!z\!>\simeq2.3$; dashed line) from \citet{Chapman05}.  This shows that we are probing a wide range in optical luminosities (and potentially a wide range in $M_\mathrm{BH}$) at the epoch where the QSO and SMG populations peak.  
}
\label{fig:selection}
\end{figure}

\section{Observations, Reduction and Redshifts}\label{obsdr}

\subsection{Sample Selection}
The objective of this study is to test the link between SMGs and QSOs at the era where these populations peaked.  We have therefore selected radio-quiet QSOs with a redshift distribution matched to SMGs ($\left\langle z\right\rangle\,{\sim}\,2.3$, \citealt{Chapman05}) and with $850\,\mathrm{\mu m}$ fluxes $\gsim5$\,mJy.  The $S_{850\,\mathrm{\mu m}}\gsim5$\,mJy flux limit selects comparatively bright QSOs and SMGs and is imposed so that the sources would be sufficiently bright to be detected in CO.  We choose submm-detected QSOs from three main samples:  
we select five QSOs from a submm photometry survey of optically-bright QSOs by \citet{Omont03} -- HS1002+4400, J140955.5+562827, J154359.3+535903, HS1611+4719, and J164914.9+530316, we select two QSOs from the submm photometry survey of X-ray absorbed QSOs by \citet{Stevens05} -- RXJ121803.82+470854.6 and RXJ124913.86--055906.2, and finally we include two QSOs found in blank-field submm surveys by \citet{Chapman05} and \citet{Greve04}, SMM\,J131222.35+423814.1 and MM\,J163655+4059, respectively, and SMM\,J123716.01+620323.3, an optically-selected QSO targetted in submm photometry mode by \citet{Chapman05}.  

The final sample comprises 10 submm-detected QSOs 
with similar fluxes to the SMG sample (\citealt{Greve05}; $\gsim5$\,mJy at 850\,$\mu$m, 
assuming 850\,$\mu$m/1.2\,mm
colours typical of SMGs; \citealt{Greve04}), with a median of 10\,mJy.
The SEDs of these QSOs demonstrate that the bulk of the mm/submm
luminosity arises from dust emission and is very likely associated with
star formation, rather than an AGN (\citealt{Beelen06}; \citealt{Omont03}).
Our sample spans six orders of magnitude in QSO $R$-band brightness, 
enabling us to potentially trace how CO properties depend on QSO luminosity or $M_\mathrm{BH}$ (see Fig.~\ref{fig:selection}).

\begin{table*}
\begin{minipage}{1.0\textwidth}

\scriptsize
\caption{Summary of the new near-infrared spectroscopy and mm-wave PdBI CO observations of submm-detected $2<z<3$ QSOs, including nine new CO observations and one published object from \citet{Beelen04}.  The CO $t_\mathrm{exp}$ corresponds to the on-source integration time with the equivalent of a 6-element array.}
\label{tab:obs}
\hspace{-0.2in}
\begin{tabular}{lcccccccc}
\hline
\multicolumn{1}{l}{Source} & \multicolumn{2}{c}{Near-IR Observations} & \multicolumn{5}{c}{CO Observations$^{a}$}\\
\hline
 & \multicolumn{1}{c}{Date(s)} & \multicolumn{1}{c}{$t_\mathrm{exp}$} & \multicolumn{1}{c}{Dates} & \multicolumn{1}{l}{$t_\mathrm{exp}$} & \multicolumn{1}{c}{Noise per channel$^{f}$} & \multicolumn{1}{c}{Beam FWHM} & \multicolumn{1}{l}{Detected?} \\
 & \multicolumn{1}{c}{} & \multicolumn{1}{c}{(ks)} & \multicolumn{1}{c}{} & \multicolumn{1}{l}{(hrs)}  & \multicolumn{1}{c}{(mJy\,beam$^{-1}$)} & \multicolumn{1}{c}{} & \multicolumn{1}{l}{} \\
\hline
HS1002+4400 & 2006 Mar 08$^{b}$ & 1.8 & 2006 Sep 16,17,19,21,22 & 7.2 & 1.4 & $4.6''\times3.3''$ & Y \\
RXJ121803.82+470854.6 & -- & -- & 2001 Dec 20, 22, 24 & 9.0 & 1.2 & $6.7''\times3.2''$ & N \\
SMM\,J123716.01+620323.3 & 2005 Feb 16$^{d}$ & 4.0 & 2007 May 12,18,30, Jun 3,5,18 & 16.0 & 0.5 & $4.7''\times4.1''$ & N \\
RXJ124913.86--055906.2 & 2006 Feb 25$^{b}$ & 4.8 & 2001 Aug 5, 2002 Apr 13,14,25, May 6 & 9.4 & 1.8 & $8.8''\times6.6''$ & Y \\
SMM\,J131222.35+423814.1 & 2004 Jun 25$^{d}$ & 2.0 & 2007 Jun 2,4,9,10,22,23 & 9.3 & 0.5 & $5.2''\times4.1''$ & Y \\
J140955.5+562827 & 2006 Mar 06$^{b}$ & 1.2 & 2002 May, June, 2003 Jan-Mar$^{e}$ & 37.0 & 0.6 & $2.4''\times1.7''$ & Y \\
J154359.3+535903 & 2006 Feb 25$^{b}$ & 4.8 & 2006 Jun 4,5,9,11,19,26 & 12.1 & 1.2 & $5.6''\times3.5''$ & Y \\
HS1611+4719 & 2006 Aug 17$^{b}$ & 3.6 & 2006 Jun 19,21,27, Jul 7 & 6.3 & 2.1 & $5.0''\times4.1''$ & Y \\
MM\,J163655+4059 & 2003 Aug 03$^{c}$, 2004 Apr 07$^{d}$ & 2.4 & 2007 Apr 30, May 21 & 9.5 & 0.4 & $5.3''\times4.4''$ & N \\
J164914.9+530316 & 2006 Mar 06$^{b}$ & 1.8 & 2006 Jul 20,23, Aug 15,18 & 6.5 & 1.6 & $5.5''\times3.7''$ & N\\
\hline
Notes: \\
\end{tabular}

(a) All observed in the CO(3--2) transition, except for RXJ121803.82+470854.6 observed in CO(2--1)\\
(b) From UIST spectroscopy (this work) \\
(c) From \citet{Swinbank04}\\
(d) From \citet{Takata07} and \citet{Alexander07}\\
(e) From \citet{Beelen04} \\
(f) With a channel width of 20\,MHz\\
\end{minipage}
\end{table*}
\normalsize

\subsection{Near-Infrared Spectroscopy}\label{nirspec}

The QSOs in this sample were spectroscopically identified through
rest-frame UV spectroscopy.  However, since the rest-frame UV emission
lines can be extremely broad (FWHM$\sim5000\,\mathrm{km\,s^{-1}}$), and can be offset in velocity
with respect to the systemic redshift, precise systemic redshifts are
required to ensure the CO emission falls within the PdBI receiver bandwidth.  
To obtain precise systemic redshifts for our targets we use the 
nebular emission
lines of [O{\sc iii}]$\lambda$5007 and H$\beta$ which 
are redshifted to near-infrared wavelengths.  

Seven of our submm-detected QSOs were observed with the United Kingdom InfraRed Telescope 
(UKIRT) 1--5\,$\mu$m Imager Spectrometer (UIST; \citealt{Hasinger04}).  
We used the $HK$ grism which provides a spectral
resolution of $\lambda / \Delta \lambda \sim 1000$ and covers a wavelength
range 1.4--2.4$\mu$m, thus allowing us to cover H$\beta$, [O{\sc
  iii}]$\lambda$4959,5007 and H$\alpha$ simultaneously 
for most of our QSOs.  The QSOs were observed between 2006 Feb and 2006 Oct.  All
observations were taken in $\lsim0.8''$ seeing and clear conditions, and were 
carried out using the standard ABBA configuration where 
the QSO was nodded by 6--12$''$ along the slit to achieve sky
subtraction.  Individual exposures were 240\,s and the total
integration times varied between 1.2\,ks and 4.8\,ks depending 
on the $K$-band magnitude of the QSO (see Table~\ref{tab:obs}).
In addition, three QSOs in our sample have suitable archival spectroscopy from 
\citet{Takata07} and \citet{Swinbank04, Swinbank06}.  

\begin{figure*}
\epsfig{file=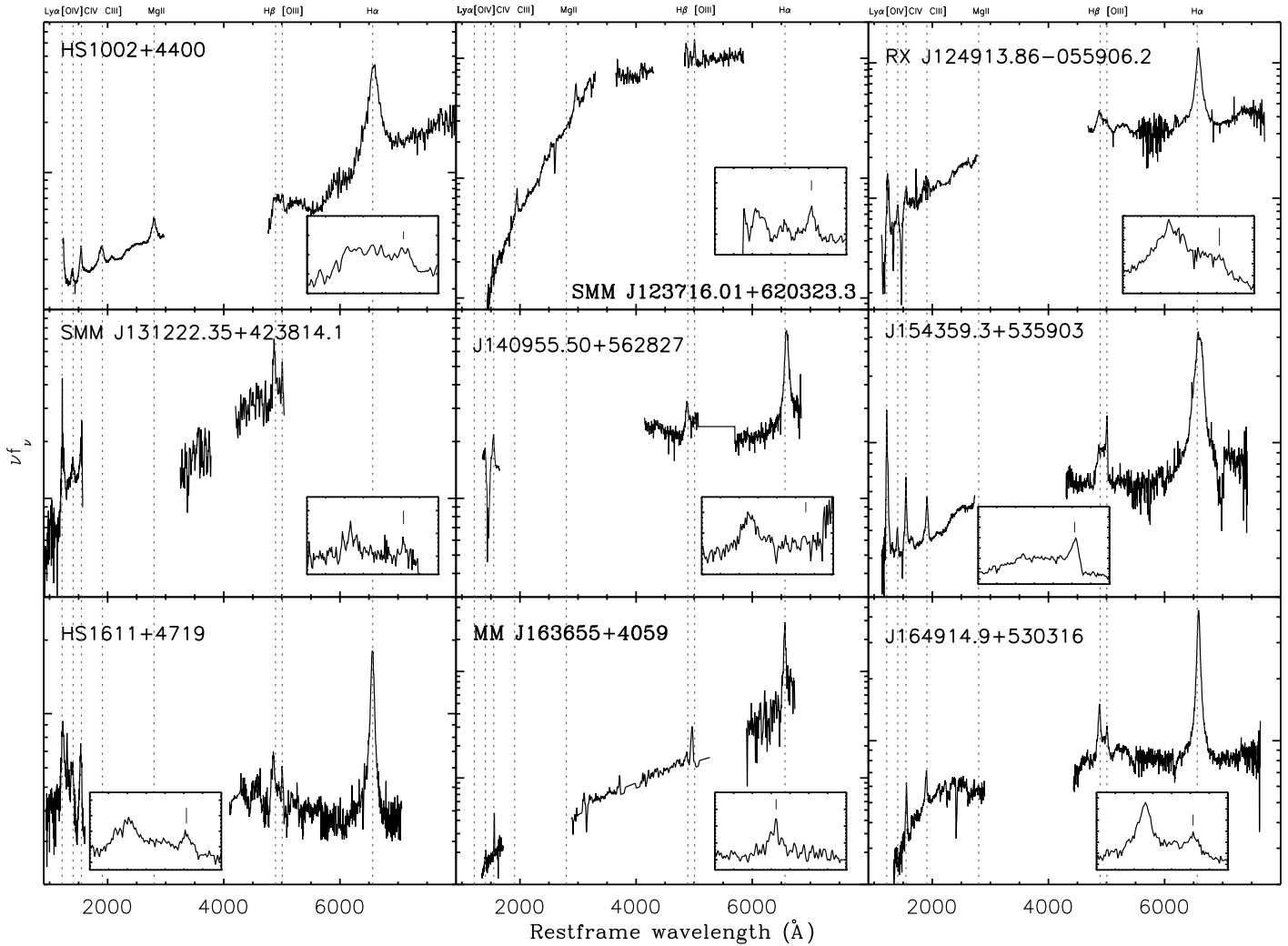,width=0.80\textwidth,angle=90}
\caption{Restframe UV--near-infrared spectra for our sample of submm-detected QSOs.  The insets show the region around the H$\beta$ and [O{\sc iii}]$\lambda$5007 emission lines (all insets cover the wavelength range 4675--$\mathrm{5100\AA}$), except in the case of J163655, and the redshifts derived from the [O{\sc iii}]$\lambda$5007 emission are indicated.  The redshift for J163655 was taken from the redshifted H$\alpha$ due to ambiguity in the [O{\sc iii}]$\lambda$5007 emission lying near the edge of the wavelength coverage of the spectrograph; the inset therefore shows the region around H$\alpha$. The region around 5100--$\mathrm{5500\AA}$ in J1409 has been masked due to strong atmospheric absorption.}
\label{fig:nirspec}
\end{figure*}

The relevant {\sc orac-dr} pipeline
\citep{Cavanagh03} was used to sky-subtract, extract, wavelength
calibrate, flat-field and flux calibrate the data (and in the case
of integral field unit observations, forms the datacube). 
Spectra for the QSOs in our sample are shown in Fig.~\ref{fig:nirspec}.  To
derive systemic redshifts, we
simultaneously fit the H$\beta$ and [O{\sc iii}]$\lambda$4959,5007 emission lines with Gaussian profiles.
The emission line flux ratio of the [O{\sc iii}]$\lambda$4959,5007 emission lines is fixed such that $I_{4959}/I_{5007}$=2.9, 
and the [O{\sc iii}]$\lambda$4959,5007 line widths are assumed to be the same.  
The fit assumes a linear continuum 
plus Gaussian emission line profiles allowing the velocity 
centroid, flux and width of the H$\beta$ line profile to be different from [O{\sc iii}].  
This makes a total of seven parameters to be fitted.
We also derive emission line fluxes and widths for the H$\alpha$
emission line (which are important for deriving $M_\mathrm{BH}$ using the method of \citealt{Greene05}) 
in a similar manner, fitting a single Gaussian emission
line profile superimposed on a linear continuum.  The resulting redshifts
and fluxes are given in Table~\ref{tab:nir}.  These redshifts were subsequently used to 
target the CO emission with PdBI (see \S \ref{mmdata}).

Due to the very weak [O{\sc iii}]$\lambda$5007 emission in RXJ124913.86, it 
was very difficult to measure a reliable redshift for the CO observations, and 
thus a range of possible redshifts were covered on this QSO ($z\sim2.24$--2.26).

During the course of assembling the data for this project, we obtained
a CO detection of HS1611 after securing a systemic redshift from
identification of [O{\sc iii}] with UIST.  The positional centroid of
the CO was offset by $\sim$1.3\,arcsec north of the optical and
near-infrared QSO position, possibly indicating a companion, 
and HS1611 was re-observed with the UIST
integral field unit (IFU) for a total of 3.6\,ks on 2006 Sep 01.  
The UIST IFU uses an image slicer to take a
$3.3''\times6.0''$ field and divides it into 14 slices of width
$0.24''$.  The dispersed spectra from the slices are reformatted on the
detector to provide two-dimensional spectro-imaging, in our case also
using the $HK$ grism.  We reduced the data using the relevant {\sc
orac-dr} pipeline which extracts, flatfields, wavelength calibrates the
data and forms the datacube.  We discuss these data in \S \ref{qc25}.

\subsection{Millimetre Interferometry}\label{mmdata}

We surveyed seven QSOs for CO emission using IRAM PdBI between 2006 June to 2007 June 
and two QSOs between 2001 December and 2002 April in conditions with 
good atmospheric phase stability and reasonable transparency. The observations were carried out using 
3, 4, 5, and 6 antennae in the D configuration, giving a total of 3, 6, 10, and 15 baselines, respectively.  
For uniformity with the previous observations observations of SMGs, which will comprise our comparison sample, 
we have aimed to achieve the same 1-$\sigma$ sensitivity as for the SMG survey \citep{Greve05} for the new 
CO observations of our sample of submm-detected QSOs: $S_\mathrm{CO}\,\Delta\,v\sim 0.3$\,Jy\,km\,s$^{-1}$. 
We also include results on a single QSO from the literature, J140955.5 from \citet{Beelen04}, 
to yield a final sample of 10 QSOs. The PdBI observations are summarised in Table~\ref{tab:obs}.

For data taken prior to 2007 January, the spectral correlator was 
adjusted to detect the line with a frequency resolution of 2.5\,MHz at a bandwidth of 580\,MHz. 
For data taken after January 2007, the spectral correlator was adjusted to detect the
line with a frequency resolution of 2.5\,MHz in the 1\,GHz band of
the new generation receivers. The visibilities were resampled to a frequency resolution of 5\,MHz.

The data were calibrated, mapped and analyzed in the IRAM 
{\sc gildas}\footnote{http://www.iram.fr/IRAMFR/GILDAS} software
package \citep{GuilloteauLucas00}.  Anomalous and high phase-noise visibilities were flagged.  
Passband calibrations were typically made using a nearby bright quasar.  
Phase and amplitude calibration within each observation were calibrated using 
frequent observations of nearby quasars approximately every 20\,min.  
The flux calibration was done using standard calibrators, including
3C\,273, MWC\,349, 1637+574, 1749+096, 0923+392, 1243-072, and CRL618, for example.
We estimate a conservative error on the flux calibration of 10--15\%.

For each QSO the central observing frequency of the 3.1\,mm band receiver was tuned 
to either the redshifted CO(3--2) or CO(2--1) rotational transition, depending on the systemic 
near-infrared redshift derived from our spectroscopy given in Table~\ref{tab:nir}.  
In the case of RXJ124913.86, the uncertainty on the redshift was large enough that we performed two 
separate observations with different central frequency tunings of 106.160 and 106.720\,GHz to ensure that the CO(3--2) line was covered.
We have combined these data to cover more effective bandwidth and have centered the resulting spectrum 
at the mean central observed frequency, 106.435\,GHz, corresponding to CO(3--2) redshifted to $z\simeq2.2487$.

Naturally weighted datacubes were made and inspected for line emission close to
the QSO position.  Fig.~\ref{fig:maps} shows the velocity-integrated mm emission 
maps over the line in each field.
In Fig.~\ref{fig:lines}, we show the spectra of the CO emission
in the brightest pixel of the central source in each velocity-integrated mm emission map, centered on the 
systemic redshift of each system.  We have included the maps and spectra of 
non-detections 
and in the latter cases we show the maps averaged over the 
central 500\,km\,s$^{-1}$ of the spectra (the average line width of our sample)
and spectra extracted at the 
phase tracking centre (or at the continuum peak for J164914.9).
In all cases, we ignore the small correction 
for the primary beam attenuation, given that the detected sources are all 
close to the phase centres of the maps.  

\section{Analysis and Results}\label{results}

\subsection{CO properties}

We present new CO data for nine QSOs, including five new CO detections. 
Four of our QSOs have CO detections with S/N$\gsim5$, and an additional QSO is marginally 
detected in CO with S/N$\simeq2.5$ with emission coincident with the phase centre (see Fig.~\ref{fig:lines}).
None of the QSOs appear to be resolved with the $\simeq6$\,arcsec beams ($\sim50$\,kpc FWHM).  
We include a previous CO detection reported by \citet{Beelen04} in our sample of submm-detected QSOs.
In total therefore this gives six detections in our sample of 10 QSOs, 
yielding a detection rate comparable to that for SMGs by \citet{Greve05} of $\sim60$\%.

Looking at the non-detections, it is unlikely that the cause of the failure to detect 
CO emission in these four QSOs is due to 
the systemic redshifts being wrong, as the quality of the optical/near-infrared spectral data 
for these is high and similar to the successful detections (see Fig.~\ref{fig:nirspec}). 
Similarly, it is unlikely that the velocity offset of the CO emission and the systemic redshift
are larger than the correlator coverage (the rms deviation of 
$z_\mathrm{NIR}$ and $z_\mathrm{CO}$ for our CO-detected QSOs is 0.0046), since only a few systems detected at 
high-redshift so far show larger offsets (e.g.~\citealt{Hainline04}). 
It is also unlikely that the CO line widths are $\gg1000\,\mathrm{km\,s^{-1}}$ since our 
typical detected lines are $\lesssim550\,\mathrm{km\,s^{-1}}$ which is much less than the correlator bandwidth.  
Although extreme CO line widths could be one reason for some of the non-detections, 
the most likely reason appears to be that the QSOs are 
simply too faint in CO to be detected given the depth of our observations.  
However, the non-detections do still provide 
useful upper limits on the CO luminosity and gas masses and therefore we include 
them in our analysis.

Other than one QSO (J164914.9; \S \ref{qd25}), we do not detect the continuum at 3.1\,mm in any of the QSOs.
This is consistent with the measured submm/mm fluxes and greybody dust emission 
with a spectral form $S_\nu\propto \nu^{2+\beta}$ in the Rayleigh-Jeans regime of the spectral 
energy distribution with a dust emissivity $\beta=1.5$.  

\begin{table*}
\begin{minipage}{1.0\textwidth}
\scriptsize
\caption{Summary of the near-infrared observed properties of the submm-detected QSOs.  
}
\label{tab:nir}
\hspace{-0.2in}
\begin{tabular}{lcrcccccc}
\hline
\multicolumn{1}{l}{Source} & \multicolumn{2}{c}{Near-IR position (J2000)} & \multicolumn{1}{c}{$R$} & \multicolumn{1}{c}{$K$} & \multicolumn{1}{c}{$z_\mathrm{NIR}^{a}$} & \multicolumn{1}{c}{FWHM$_\mathrm{H\alpha}$} & \multicolumn{1}{c}{$S_\mathrm{H\alpha}$} & \multicolumn{1}{c}{FWHM$_\mathrm{[O\small{III}]}$} \\
\multicolumn{1}{c}{} & \multicolumn{1}{c}{R.A.} & \multicolumn{1}{c}{Dec.} & \multicolumn{2}{c}{(Vega)}  & \multicolumn{1}{c}{} & \multicolumn{1}{c}{(km\,s$^{-1}$)} & \multicolumn{1}{c}{($\times 10^{-19}\,$W/m$^{2}$)} & \multicolumn{1}{c}{(km\,s$^{-1}$)} \\
\hline
HS1002+4400 & 10\,05\,17.43 & 43\,46\,09.3 & 16.1 & 14.2 & $2.1015\pm0.0010$ & $10000\pm1000$ & $750\pm100$ & $1500\pm300$ \\
RXJ121803.82+470854.6 & 12\,18\,04.54 & 47\,08\,51.0 & 19.5 & 17.0 & 1.7416 & -- & -- & -- \\
SMM\,J123716.01+620323.3 & 12\,37\,16.00 & 62\,03\,23.4 & 20.2 & 15.6 & $2.0568\pm0.0013$ & $2100\pm500^{b}$ & $48\pm10^{b}$ & $1100\pm200$ \\
RXJ124913.86--055906.2 & 12\,49\,13.85 & $-05$\,59\,19.4 & 16.2 & 13.5 & $2.2400\pm0.0100$ & $4820\pm200$ & $1000\pm100$ & $700\pm300$ \\
SMM\,J131222.35+423814.1 & 13\,12\,22.32 & 42\,38\,13.9 & 20.2 & 18.0 & $2.5543\pm0.0010$ & $2600\pm1000^{b}$ & $60\pm10^{b}$ & $670\pm300$ \\
J140955.5+562827 & 14\,09\,55.50 & 56\,28\,27.0 & 17.0 & 14.9 & $2.5758\pm0.0050$ & $4250\pm200$ & $220\pm20$ & $600\pm200$ \\
J154359.3+535903 & 15\,43\,59.44 & 53\,59\,03.2 & 16.8 & 14.3 & $2.3692\pm0.0015$ & $8280\pm300$ & $770\pm50$ & $1000\pm200$ \\
HS1611+4719 & 16\,12\,39.90 & 47\,11\,57.0 & 17.1 & 15.1 & $2.4030\pm0.0008$ & $4010\pm300$ & $300\pm20$ & $1000\pm300$ \\
MM\,J163655+4059 & 16\,36\,55.79 & 40\,59\,10.5 & 23.2 & 19.1 & $2.6070\pm0.0006$ & $3000\pm400$ & $16\pm2$ & $1840\pm400$ \\
J164914.9+530316 & 16\,49\,14.90 & 53\,03\,16.0 & 16.9 & 14.2 & $2.2704\pm0.0009$ & $3100\pm300$ & $580\pm50$ & $1000\pm200$ \\
\hline
Notes:\\
\end{tabular}

(a) All redshifts are derived from [O{\sc iii}], except for RXJ121803.82, where the redshift is derived from C{\sc iv}, C{\sc iii}] and Mg{\sc ii} by \citet{Page01} using WHT data\\
(b) H$\alpha$ flux is calculated from the H$\beta$ flux, assuming an intrinsic $S_\mathrm{H\alpha}/S_\mathrm{H\beta}\simeq3.1$ \citep{Alexander07}, and the given line width is for H$\beta$ \\
\end{minipage}
\end{table*}
\normalsize

\begin{table*}
\begin{minipage}{1.0\textwidth}
\scriptsize
\caption{Summary of the mm and CO observed properties of the submm-detected QSOs.}
\label{tab:co}
\hspace{-0.2in}
\begin{tabular}{lcrccccccc}
\hline
\multicolumn{1}{l}{Source} & \multicolumn{2}{c}{CO position (J2000)} & \multicolumn{1}{c}{Offset} & \multicolumn{1}{c}{Resolution$^{g}$} & \multicolumn{1}{c}{$z_\mathrm{CO}^{h}$} & \multicolumn{1}{c}{$S_\mathrm{CO}\,\Delta\,v^{h}$} & \multicolumn{1}{c}{FWHM$_\mathrm{CO}^{h}$} & \multicolumn{1}{c}{$850\,\mathrm{\mu m}$ flux} & \multicolumn{1}{c}{1.2\,mm flux} \\
\multicolumn{1}{c}{} & \multicolumn{1}{c}{RA} & \multicolumn{1}{c}{Dec.} & \multicolumn{1}{c}{(arcsec)} & \multicolumn{1}{c}{(kpc)} & \multicolumn{1}{c}{} & \multicolumn{1}{c}{(Jy\,km\,s$^{-1}$)} &  \multicolumn{1}{c}{(km\,s$^{-1}$)} & \multicolumn{1}{c}{(mJy)} & \multicolumn{1}{c}{(mJy)} \\
\hline
HS1002+4400 & 10\,05\,17.43 & 43\,46\,10.1 & 0.5 & 33 & $2.1015\pm0.0007$ & $1.7\pm0.3$ & $640\pm160$ & -- & $4.2\pm{0.8}^{a}$ \\
RXJ121803.82+470854.6  & -- & -- & -- & 42 & -- & $\leq0.6$ & -- & $6.8\pm{1.2}^{b}$ & -- \\
SMM\,J123716.01+620323.3  & -- & -- & -- & 37 & -- & $\leq0.3$  & -- & $5.3\pm{1.7}^{c}$ & -- \\
RXJ124913.86--055906.2  & 12\,49\,13.91 & $-05$\,59\,20.1 & 0.6 & 125 & $2.2470\pm0.0016$ & $1.3\pm0.4$ & $1090\pm340$ & $7.2\pm{1.4}^{b}$  & -- \\
SMM\,J131222.35+423814.1 & 13\,12\,22.23 & 42\,38\,13.9 & 0.9 & 38 & $2.5564\pm0.0011$ & $0.4\pm0.1$ & $550\pm220$ & $3.0\pm{0.9}^{c}$ & -- \\
J140955.5+5628$27^{e}$ & 14\,09\,55.50 & 56\,28\,26.4 & -- & 17 & $2.5832\pm0.0001$ & $2.3\pm0.2$ & $310\pm30$ & -- & $10.7\pm{0.6}^{a}$ \\
J154359.3+535903 & 15\,43\,59.43 & 53\,59\,03.4 & 0.7 & 38 & $2.3698\pm0.0006$ & $1.0\pm0.2$ & $520\pm140$ & -- & $3.8\pm{1.1}^{a}$ \\
HS1611+4719  & 16\,12\,39.90 & 47\,11\,58.3 & 0.6 & 37 & $2.3961\pm0.0002$ & $1.7\pm0.3$ & $230\pm40$ & -- & $4.6\pm{0.7}^{a}$ \\
MM\,J163655+4059  & -- & -- & -- & 39 & -- & $\leq0.2$ & -- & --  & $2.2\pm{0.6}^{d}$ \\
J164914.9+5303$16^{f}$ & -- & -- & -- & 39 & -- & $\leq0.8$ & -- & -- & $4.6\pm{0.8}^{a}$ \\
\hline
Notes: \\
\end{tabular}

(a) 1.2\,mm flux from \citet{Omont03} \\
(b) 850\,$\mu$m flux from \citet{Page01}\\
(c) 850\,$\mu$m flux from \citet{Chapman05}\\
(d) 1.2\,mm flux from \citet{Greve04} \\
(e) CO parameters from \citet{Beelen04}\\
(f) Detected in continuum only\\
(g) Approximate CO Resolution at the target redshifts, given the beamsizes in Table~\ref{tab:obs}\\
(h) This quantity was obtained by fitting a Gaussian distribution to the CO spectrum (see text)\\
\end{minipage}
\end{table*}
\normalsize

The luminosity, velocity width and spatial extent of the CO line
emission can be used to place limits on the gas and dynamical mass of
each system.  For each CO-detected QSO a Gaussian profile is fit to either 20- or 30-MHz 
binned data by minimising the $\chi^{2}$ statistic.  
The $1\,\sigma$ errors on the best-fitting parameters are quoted where appropriate.  To calculate the 
velocity-integrated flux density of the CO emission, we integrate the best-fitting Gaussian profile 
(note that we do not fit the line and continuum simultaneously).  The 
error on the velocity-integrated flux density is calculated by bootstrapping from a Normal distribution 
centered on the best fitting amplitude above, with a width given by the $1\,\sigma$ error in the fitted amplitude, 
while holding the FWHM and central velocity of the best-fitting Gaussian profile fixed.  The noise in the 
averaged channel maps, constructed from the average emission over the full width at zero intensity (FWZI) in each QSO, 
is determined by summing over the weights and takes into account the phase noise visibilities.

For the non-detections we calculate 3\,$\sigma$ upper limits 
$S_\mathrm{CO}=3\,\sigma\,(\delta v\,\Delta v_\mathrm{FWHM})^{1/2}$, 
where $\sigma$ is the channel-to-channel rms noise, $\delta v$ is the velocity resolution and 
$\Delta v_\mathrm{FWHM}$ is the line width (see e.g.\ \citealt{Greve05}; \citealt{SeaquistIvisonHall95}).  
We use the channel-to-channel rms noise of the spectra binned to a resolution of 100\,MHz 
and adopt a line width of $500\,\mathrm{km\,s^{-1}}$.  Since our detections have typical line widths $\lesssim550\mathrm{km\,s^{-1}}$, 
this is a conservative assumption.  In addition, this line width assumption 
facilitates the direct comparison of our non-detections with those of SMGs in \citet{Greve05}.

We have summarised the line properties of each source (the CO position, $z_\mathrm{CO}$, $L'_\mathrm{CO(3-2)}$ and FWHM) in Tables~\ref{tab:co} and \ref{tab:derived}.  Offsets of the CO emission from the radio position are given in Table~\ref{tab:co}, assuming that the source is unresolved using equation B2 in \citet{Ivison07}, and are less than $2\,\sigma$ significant in most cases.  We note that we discover serendipitous $\gsim4\,\sigma$ sources in two of the fields surveyed which may be associated with gas-rich companions to the QSOs (SMM\,J123716.01 and SMM\,J131222.35).  Overdensities of submm sources have already been found by \citet{Stevens04} around similar QSOs on slightly bigger scales than those probed here.  We now discuss the CO properties of each QSO detection and non-detection in detail, as well as 
any serendipitous detections.  

\subsubsection{HS1002+4400}

The integrated CO emission in this QSO is detected at $\simeq6\,\sigma$, 
 $0.8''$ north of the Sloan Digital Sky Survey (SDSS; \citealt{Fan99}) position  (see Fig.~\ref{fig:maps}).  The CO(3--2) spectrum of HS1002+4400 has been binned to a frequency resolution 
of 20\,MHz (54\,km\,s$^{-1}$) and is shown in Fig.~\ref{fig:lines}.  
 The CO emission line is well fitted by a Gaussian profile with a 
FWHM of 640$\pm$160\,km\,s$^{-1}$, a velocity-integrated flux density of 
$S_\mathrm{CO(3-2)}=1.7\pm{0.3}\,\mathrm{Jy\,km\,s^{-1}}$, and a 
CO redshift of $z_\mathrm{CO}=2.1015\pm0.0007$, offset by $20\pm65$\,km\,s$^{-1}$ from the near-infrared systemic redshift.
No significant continuum emission is detected from the line-free region 
down to a 1-$\sigma$ limit of 0.4\,mJy.

\subsubsection{RXJ121803.82+470854.6}

Neither CO(2--1) emission at the systemic redshift nor continuum emission are seen in the map near the QSO position.  
To calculate a 3\,$\sigma$ upper limit to the velocity-integrated CO(2--1) line flux of RXJ121803.82, we extract a spectrum at the 
QSO position (Table~\ref{tab:nir}) and rebin the spectrum to a frequency resolution of 100\,MHz (360\,km\,s$^{-1}$), resulting in a 
channel rms of 0.5\,mJy, and so derive an upper limit of 0.6\,Jy\,km\,s$^{-1}$, assuming a line width of 500\,km\,s$^{-1}$, centred on 
the $z_\mathrm{NIR}=1.7416$. No significant continuum emission is detected 
down to a 1-$\sigma$ limit of 0.2\,mJy.

We do find two $\simeq4\,\sigma$ serendipitous detections in the datacube, offset in velocity from the central frequency tuning, and approximately 10-20\,arcsec away from the phase centre. Neither show strong broad line detections in the spectra extracted at the peak locations, so we believe these are unlikely to be true sources.

\subsubsection{SMM\,J123716.01+620323.3}\label{J1237}

Neither CO(3--2) emission nor continuum emission are seen in the datacube near the QSO position. To calculate a 3\,$\sigma$ upper limit to the velocity-integrated CO(3--2) line flux, we extract a spectrum at the radio position (Table~\ref{tab:nir}) and rebin the spectrum to a frequency resolution of 100\,MHz (270\,km\,s$^{-1}$), resulting in a channel rms of 0.2\,mJy, giving an upper limit of 0.3\,Jy\,km\,s$^{-1}$, assuming a line width of 500\,km\,s$^{-1}$ at the systemic redshift from [O{\sc iii}]$\lambda$5007 (Table~\ref{tab:nir}). No significant continuum emission is detected down to a 1-$\sigma$ limit of 0.2\,mJy.

We find a $\simeq\!\!\!\!4\,\sigma$ detection in the datacube, offset in velocity from the central frequency tuning and approximately 18\,arcsec\footnote{Note that the map noise at this distance from the phase centre is only slightly degraded since the primary half power beam width (HPBW) at this frequency is $\simeq45$\,arcsec.} south of the nominal pointing position at 12\,37\,15.36, 62\,03\,06.5 (J2000), indicating a redshift of $z_\mathrm{CO}=2.0597\pm0.0001$, which is close to the QSO $z_\mathrm{NIR}$, suggesting it may be a gas-rich companion of the QSO with a separation of 170\,kpc.  See Fig.~\ref{fig:extras}.  This position corresponds to within 0.5\,arcsec of an $I=24.7$ counterpart \citep{Smail04} with a radio flux of $\leq15\,\mu$\,Jy \citep{Biggs07} which is also undetected in the \textit{Chandra} observations \citep{Alexander03}.  The CO line has an integrated flux of $S_\mathrm{CO(3-2)}=0.2\pm{0.06}\,\mathrm{Jy\,km\,s^{-1}}$ and is very narrow (FWHM=$100\pm30\,\mathrm{km\,s^{-1}}$).  

\subsubsection{RXJ124913.86--055906.2}

The integrated CO emission is detected at $\simeq5\,\sigma$, 
1.1\,arcsec southeast of the QSO position (Table~\ref{tab:nir}).
The CO(3--2) spectrum of RXJ124913.86 has been binned to a frequency resolution 
of 30\,MHz (84\,km\,s$^{-1}$) and is shown in Fig.~\ref{fig:lines}. We have combined two sets of data with different central frequency tunings (see \S \ref{mmdata}) and have arbitrarily set the central frequency to 106.435\,GHz, corresponding to CO(3--2) redshifted to $z=2.2487$.  
The CO emission line is fitted by a Gaussian profile with a 
FWHM of 1090$\pm$340\,km\,s$^{-1}$, a velocity-integrated flux density of 
$S_\mathrm{CO(3-2)}=1.3\pm{0.4}\,\mathrm{Jy\,km\,s^{-1}}$, 
and a CO redshift of $z_\mathrm{CO}=2.2470\pm0.0016$, offset by $-170\pm140$\,km\,s$^{-1}$ from 
the near-infrared redshift.  This is our broadest CO line detection.
No significant continuum emission is detected from the line-free region 
down to a 1-$\sigma$ limit of 0.8\,mJy.

\subsubsection{SMM\,J131222.35+423814.1}\label{J1312}

The integrated CO emission of this QSO is marginally detected at $\simeq2.5\,\sigma$, 
coincident with the phase centre of the PdBI observations (Table~\ref{tab:co}).
The CO(3--2) spectrum of SMM\,J131222.35 has been binned to a frequency resolution 
of 20\,MHz (62\,km\,s$^{-1}$) and is shown in Fig.~\ref{fig:lines}.  
The CO emission line is fitted by a Gaussian with a 
FWHM of 550$\pm$220\,km\,s$^{-1}$, a velocity-integrated flux density of 
$S_\mathrm{CO(3-2)}=0.4\pm{0.1}\,\mathrm{Jy\,km\,s^{-1}}$ (S/N$\simeq$3), and a CO redshift of 
$z_\mathrm{CO}=2.5564\pm0.0011$ which is consistent with the near-infrared redshift within the $1\,\sigma$ errors.
No significant continuum emission is detected from the line-free region 
down to a 1-$\sigma$ limit of 0.1\,mJy.

We find a $\simeq4\,\sigma$ serendipitous emission line in the datacube at 13\,12\,23.63, 42\,38\,19.33 (J2000), approximately 15\,arcsec northeast\footnote{Note that the map noise at this distance from the phase centre is only slightly degraded since the primary half power beam width (HPBW) at this frequency is $\simeq52$\,arcsec.} of the phase centre, corresponding to a low S/N detected line with a FWHM=450$\pm150\,\mathrm{km\,s^{-1}}$, $z_\mathrm{CO}=2.5408\pm0.0007$, 
and $S_\mathrm{CO(3-2)}=0.40\pm{0.17}\,\mathrm{Jy\,km\,s^{-1}}$ (S/N$\simeq2.5$).  See Fig.~\ref{fig:extras}.  We do not find an optical counterpart for this source down to $K\sim20$ and $z\sim24$.  If real, this source is offset by 120\,kpc and $1300\,\mathrm{km\,s^{-1}}$ from the QSO.

\subsubsection{J140955.5+562827}

We briefly summarise the details for J140955.5 here since its detection is published in (\citealt{Hainline04}; \citealt{Beelen04}).  \citet{Beelen04} report a CO(3--2) emission detection at the central observed frequency (corresponding to $z_\mathrm{CO}=2.5832\pm0.0001$) that is well fitted by a Gaussian with a FWHM\,=\,311\,$\pm\,28\,\mathrm{km\,s^{-1}}$, yielding $S_\mathrm{CO(3-2)}=2.3\pm{0.2}\,\mathrm{Jy\,km\,s^{-1}}$.

\subsubsection{J154359.3+535903}

We detect the integrated CO emission of this QSO at $\simeq5\,\sigma$, 
and coincides with the phase centre of the PdBI observations (Table~\ref{tab:co}, Fig.~\ref{fig:maps}).
The CO(3--2) spectrum of J154349.3 has been binned to a frequency resolution
of 20\,MHz (58\,km\,s$^{-1}$) and is shown in Fig.~\ref{fig:lines}.  
The CO emission line is well fitted by a Gaussian with a 
FWHM of 500$\pm$130\,km\,s$^{-1}$, a velocity-integrated flux density of 
$S_\mathrm{CO(3-2)}=1.0\pm{0.2}\,\mathrm{Jy\,km\,s^{-1}}$, 
and a CO redshift of
$z_\mathrm{CO}=2.3698\pm0.0006$ consistent with the near-infrared redshift 
of the QSO within the $1\,\sigma$ errors.
No significant continuum emission is detected from the 
line-free region 
down to a 1-$\sigma$ limit of 0.4\,mJy. 

\subsubsection{HS1611+4719}\label{qc25}

The integrated CO emission in this QSO is detected at $\simeq6\,\sigma$, 
1.3\,arcsec north of the phase center of the PdBI observations (see Fig.~\ref{fig:maps}).  
The CO(3--2) spectrum of HS1611+4719 has been binned to a frequency resolution 
of 20\,MHz (59\,km\,s$^{-1}$) and is shown in Fig.~\ref{fig:lines}.  The CO emission line is well fitted by a Gaussian with a 
FWHM of 230$\pm$40\,km\,s$^{-1}$, a velocity-integrated flux density of 
$S_\mathrm{CO(3-2)}=1.7\pm{0.3}\,\mathrm{Jy\,km\,s^{-1}}$ (S/N$\simeq$6), and a CO redshift of 
$z_\mathrm{CO}=2.3961\pm0.0002$.  No significant continuum emission is detected 
from the line-free region down to a 1-$\sigma$ limit of 0.5\,mJy.

The apparent 1.3\,arcsec spatial and $\sim\!\!-600$\,km\,s$^{-1}$ velocity
offsets between the CO and near-infrared emission might be naturally explained if the CO is associated with
a companion galaxy.  To test this we search the UIST IFU datacube for a
companion galaxy to the north, but are not able to identify any strong
line emitting galaxies to a flux limit of
$\sim$2$\times$10$^{-17}$\,W\,m$^{-2}$ (corresponding to an H$\alpha$
star-formation rate of $\sim$40\,M$_{\odot}$\,yr$^{-1}$; \citealt{Kennicutt98}).  
We note that the near-infrared spectrum shows an extended
blue wing on the [O{\sc iii}] emission line which is coincident with
the redshift of the CO emission (see \S \ref{nirspec}).  Higher resolution mm 
spectroscopy and/or deeper near-infrared spectral imaging may be
required to determine whether the spatial and velocity offsets arise due 
to a companion galaxy, an outflow from the QSO, or a merger. 
This is the only example of a potentially significant offset between the CO
and near-infrared position and/or redshifts.

\subsubsection{MM\,J163655+4059}

Neither CO(3--2) emission nor continuum emission are seen in the map near the QSO position.  To calculate a 3-$\sigma$ upper limit to the velocity-integrated CO(3--2) line flux, we extract a spectrum at the radio position (Table~\ref{tab:nir}) and rebin the spectrum to a frequency resolution of 100\,MHz (310\,km\,s$^{-1}$) centred on the near-infrared redshift (Table~\ref{tab:nir}), resulting in a channel rms of 0.2\,mJy, and we find an upper limit of 0.2\,Jy\,km\,s$^{-1}$, assuming a line width of 500\,km\,s$^{-1}$.  No significant continuum emission is detected down to a 1-$\sigma$ limit of 0.2\,mJy.

\subsubsection{J164914.9+530316}\label{qd25}

\begin{figure*}
\vspace*{0.1in}
\epsfig{file=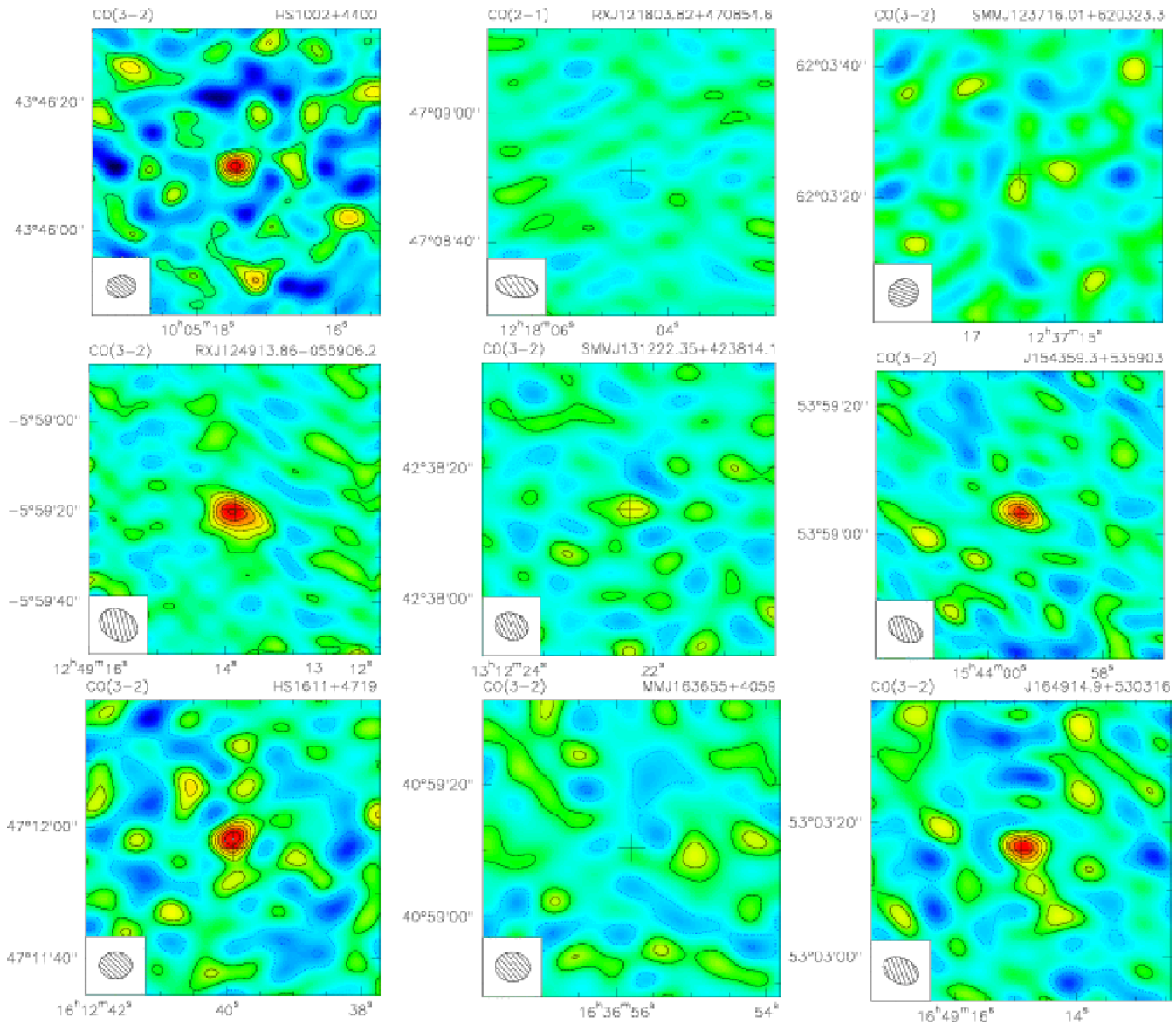,width=1.0\textwidth}
\caption{The velocity-integrated mm emission of our nine new submm-detected QSOs. The solid contours are 1, 2, 3...\,$\times\,\sigma$ (dashed lines represent equivalent negative contours), where $\sigma$ 
refers to the noise in each velocity-integrated map (integrated over the width of the CO line) determined by summing over the weights and includes a contribution from the phase noise visibilities.  These are the dirty maps (except for RXJ124913.86 which has been cleaned for presentation here to remove significant sidelobe structure) and the synthesized PdBI beams are shown in the insets as hatched ellipses for reference.  Crosses indicate the radio or optical position of each source.  \textit{Detections:}  The CO-detected sources shown here are HS1002, RXJ124913.86, J131222.35, J154359.3, and HS1611.  Note that the CO detection of J131222.35 is marginal.  J164914.9 shows a strong continuum detection but no CO detection (see \S \ref{qd25}).  The positional uncertainties are $\simeq$\,0.5, 0.6, 0.9, 0.7, 0.6 and 0.5, respectively, for the CO or continuum detected sources (ranked in RA) based on the S/N of the detections and the beam sizes (see text).  \textit{Non-detections:}  The three CO and continuum non-detections are shown for comparison and show the 3\,mm emission over the central 500\,km\,s$^{-1}$ of the spectra (the average line width of our sample).}
\label{fig:maps}
\end{figure*}

\begin{figure*}
\epsfig{file=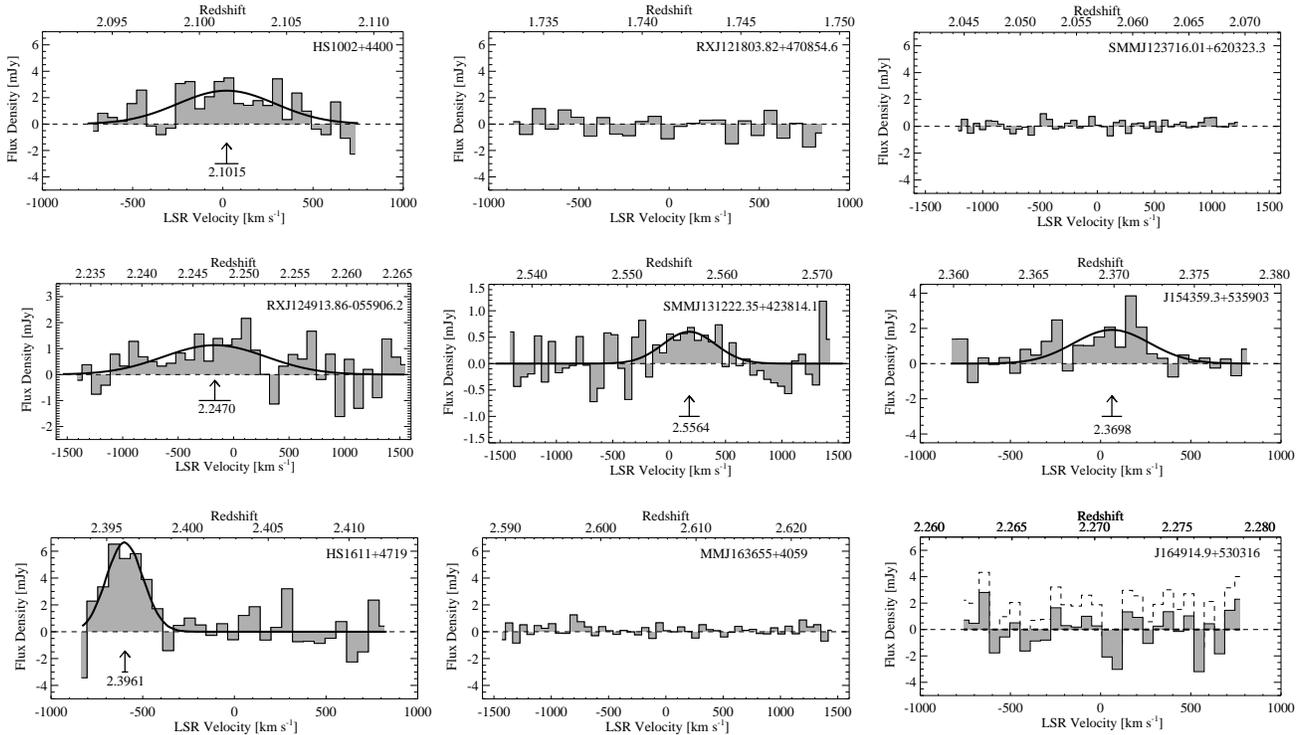,width=1.0\textwidth}
\caption{Millimetre spectra of the nine new submm-detected QSOs presented in our sample. The LSR velocity scale is relative to the near-infrared redshift (Table~\ref{tab:nir}) of each QSO, except in the RXJ124913.86, where two central frequency tunings were used, reflecting the uncertainty of the systemic redshift of the source at the time of the CO observations were made, and so we have arbitrarily centered the spectrum on CO(3--2) redshifted to $z=2.2487$.  \textit{Detections:} The CO-detected sources HS1002, J154359.3, J131222.35 and HS1611, have been binned to a frequency resolution of 20\,MHz, and RXJ124913.86 has been binned to 30\,MHz.  The best-fitting CO redshift for each QSO is obtained by fitting each spectrum with a single Gaussian distribution (overplotted) and is indicated by an arrow.  See \citet{Beelen04} for the CO spectrum of J140955.5.  \textit{Non-detections:} The CO-undetected spectra have been extracted from the phase tracking centre of the map and have been binned to a resolution of 20\,MHz for comparison with the CO detections.  The spectrum of J164914.9 has been continuum-subtracted, and the original spectrum is given as the dashed histogram.}
\label{fig:lines}
\end{figure*}

\begin{figure*}
\vspace*{0.1in}
\epsfig{file=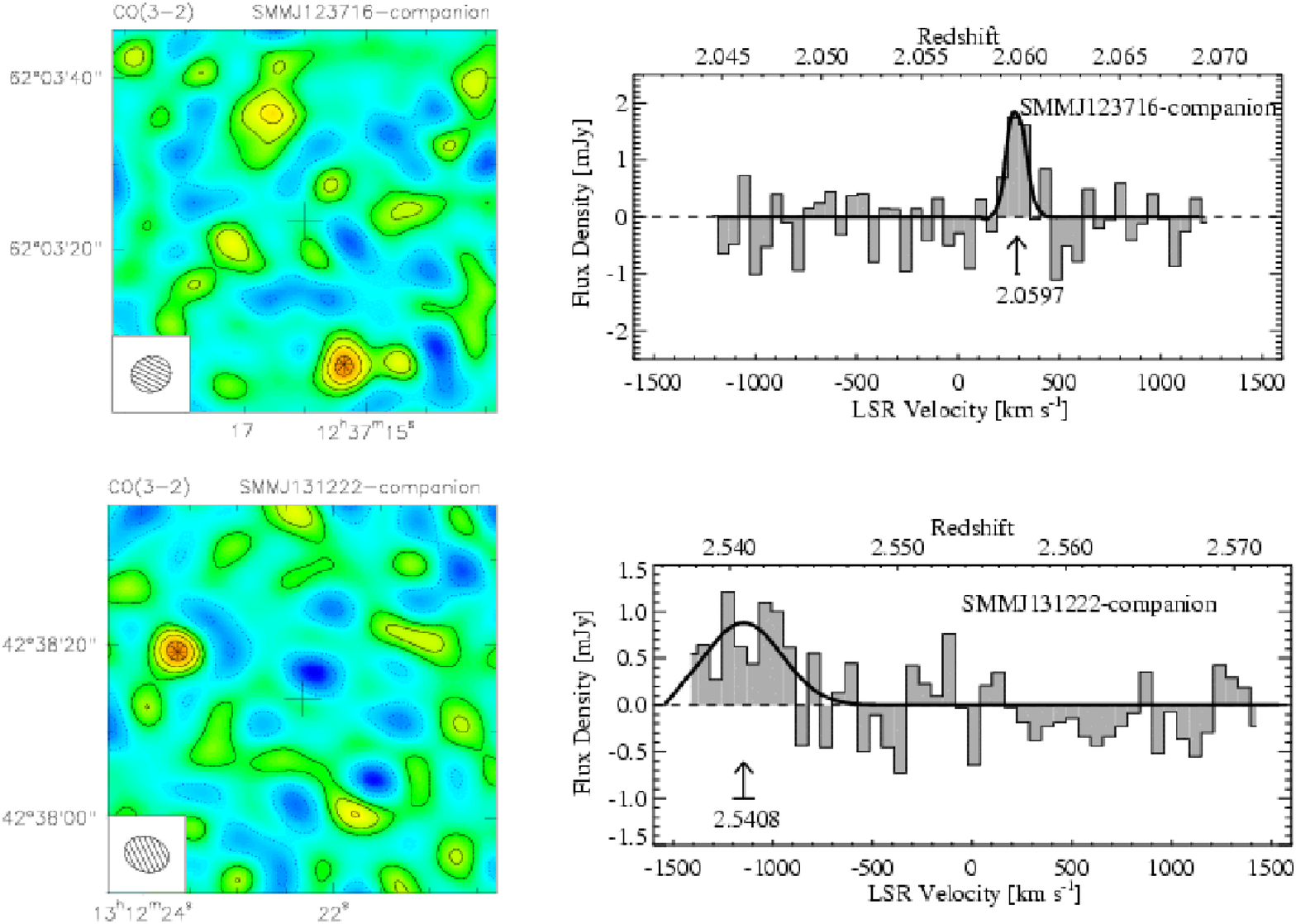,width=1.0\textwidth}
\caption{The velocity-integrated mm emission (left) and spectra (right) of serendipitous emission-line candidate 'companion' detections in two of the CO datacubes.  See \S~\ref{J1237} and \ref{J1312} for details about the individual detections.  \textit{Left:} A $\sim4\,\sigma$ source appears in each map (solid contours are 1, 2, 3...\,$\times\,\sigma$), indicated by an asterisk, offset from the PdBI phase centre (indicated by a cross).  See the caption of Fig.~\ref{fig:maps} for details.  \textit{Right:} 20-MHz binned spectra are shown for each serendipitous emission-line candidate companion source (see the caption of Fig.~\ref{fig:lines} for details).}
\label{fig:extras}
\end{figure*}

No CO(3--2) emission line is seen in this QSO, although continuum emission is detected in the averaged channel map with a flux of $1.6\pm0.28$\,mJy (S/N$\simeq$6) at a position of 16\,49\,14.853, 53\,03\,16.35 (J2000) (see Fig.~\ref{fig:maps}).  

We have determined that the continuum is dominated by synchrotron or free-free emission, with a radio spectral index, $\alpha'$, of 0.1, where $S_{\nu}\propto \nu^{\alpha'}$, based on 1.4 and 5\,GHz fluxes of $820\pm20$ and $910\pm80\,\mu\,\mathrm{Jy}$, respectively \citep{Petric06} which predicts a 3\,mm flux of 1.3\,mJy, in agreement with our detection.  Based on the 3\,mm flux, we estimate that approximately a third of the 1.2\,mm flux (1.2\,mJy of the $S_\mathrm{1.2\,mm}=4.6\pm0.8$\,mJy) is due to synchrotron/free-free emission rather than dust emission.   We calculate a 3\,$\sigma$ upper limit to the continuum-subtracted velocity-integrated CO(3--2) line flux of 0.8\,Jy\,km\,s$^{-1}$, by rebinning the spectrum to a frequency resolution of 100\,MHz (280\,km\,s$^{-1}$), resulting in a channel rms of 0.7\,mJy, and assuming a line width of 500\,km\,s$^{-1}$ centred on $z_\mathrm{NIR}=2.2704$.

\subsection{CO Luminosities and Gas Masses}

For each submm-detected QSO, we calculate the line luminosity and
estimate the total cold gas mass, $M_\mathrm{gas}$, from the integrated CO line flux following 
\citet{Solomon05}.  We assume a line luminosity ratio of 
$r_{32}=r_{21}=r=L'_\mathrm{CO(3-2)}/L'_\mathrm{CO(1-0)}=1$ 
(i.e.~a constant brightness temperature).
For our QSO sample, we derive a median of $L'_\mathrm{CO(1-0)}=(3.1\pm0.9)\times10^{10}\,\mathrm{K\,km\,s^{-1}\,pc^{2}}$ for the non-detections and a median of 
$L'_\mathrm{CO(1-0)}=(4.2\pm1.2)\times10^{10}\,\mathrm{K\,km\,s^{-1}\,pc^{2}}$ for only the CO detections (see Table~\ref{tab:co}).
We then convert the CO luminosities into total cold gas masses 
(including a correction factor for Helium) using 
$M_\mathrm{gas}=M($H$_2+$He$)=\alpha\,L'_\mathrm{CO(1-0)}$.  We use 
$\alpha=0.8\,$M$_{\odot}(\mathrm{K\,km\,s^{-1}\,pc^{2}})^{-1}$ as the CO-to-gas conversion 
factor appropriate for local galaxy populations exhibiting similar
levels of star formation activity to submm-bright galaxies or QSOs (e.g.\ ULIRGs).  
$r=1$ and $\alpha=0.8$ are also used by \citet{Greve05} for the SMG sample and so our QSOs can be 
directly compared with SMGs 
(if these values are appropriate for both populations). 
The gas masses for our QSOs can be found in Table~\ref{tab:derived}.
The median gas mass of the entire sample is $(2.5\pm0.7)\times 10^{10}$\,M$_{\odot}$ 
when the upper limits are included, and 
the median gas mass for the CO-detected QSOs is $(3.4\pm0.8)\times 10^{10}$\,M$_{\odot}$.
These gas masses are comparable to those observed in higher-redshift ($z\gsim4$) QSOs (e.g.~\citealt{Riechers06}).

Our observations are of $J\geq2$ CO transitions, and it is possible that 
significant amounts of cold, possibly subthermal molecular gas could be 
present, but only detectable in lower CO transitions.  
Moreover, if the gas is metal-poor, it is possible that the gas mass could be higher, or we could be missing clumpy 
dense gas, as these CO observations primarily trace diffuse ISM.  
Fortunately, these possibilities are unlikely to have a large effect:  \citet{Riechers06} 
show that the CO emission in three well-studied $z>4$ QSOs is well-described by a 
single centrally-concentrated molecular gas component and is highly excited.  But a concern remains that if the CO-to-gas conversion 
factor for a faint extended component is higher (e.g.\ Galactic), a higher H$_{2}$ 
mass may be hidden in such an extended component.  Nevertheless, with suitable caution these observations 
can still be directly comparable to the SMG sample as they suffer 
from the same uncertainties and potential biases.

\subsection{Line Widths and Dynamical Masses}\label{fwhm}

The median CO line width of our sample is $550\pm180\,\mathrm{km\,s^{-1}}$ (Table~\ref{tab:co}).  CO line widths can be directly converted into dynamical masses, assuming a size and inclination for the gas reservoirs.  In the following we assume a disk model \citep{Solomon05}.  Following \citet{Tacconi06}, we derive $v_{c}\,sin(i)$, where $v_c$ is the circular velocity at the outer CO radius and $i$ is the inclination of the gas disk relative to the sky plane, by dividing the FWHM of the CO line by 2.4.  We calculate the dynamical masses as $M_\mathrm{dyn}=\,R\,v_{c}^{2}\,csc^{2}(i)/G$, where R is the radius of the gas disk which we assume to be 2\,kpc, and list these in Table~\ref{tab:derived}.  With a typical resolution of $\sim40$\,kpc for our sample (see Table~\ref{tab:co}), putting a 2\,kpc radius limit is a major assumption (as are assumptions on the system inclination), although we explain why a 2\,kpc disk is probably an appropriate assumption for our QSOs in \S \ref{fwhm_discuss}.  We note that our estimates are conservative as the dynamical masses will be higher by a factor of about two if a merger model is adopted \citep{Genzel03}.  

For the CO-detected QSO sample we have a median dynamical mass of M$(<2$\,kpc$)\simeq(2.5\pm1.6)\times 10^{10}\,csc^2(i)$.  The main uncertainties in the dynamical mass limits are the assumed disk size and the inclination angle, $i$.  We note that one of our detected QSOs, RXJ124913, has a CO FWHM amongst the largest ever observed for a QSO at any redshift, although our measurement is highly uncertain (the median FWHM for all CO-detected QSOs, including lensed and $z>4$ objects, is $300\,\mathrm{km\,s^{-1}}$; \citealt{Solomon05}), suggesting that RXJ124913 is viewed at a high inclination angle or that the CO disk size is larger than we have assumed.  However, high resolution observations of the CO distribution are needed to better constrain the gas disk sizes, inclination angles, and hence dynamics of our QSOs. 

\subsection{Far-infrared Luminosities, SFRs and SFEs}\label{lfircalc}

We have derived $L_\mathrm{FIR}$ for the submm-detected QSOs from their restframe far-infrared fluxes and report these in Table~\ref{tab:derived}.  We scale a modified greybody model for far-infrared emission to match the 850 or $1200\,\mathrm{\mu m}$ photometry assuming a dust temperature of $T_\mathrm{d}=40$\,K and a dust emissivity factor of $\nu^{\beta}$, with $\beta=1.5$ (see e.g.\ \citealt{Coppin07}) and integrate the SED to obtain $L_\mathrm{FIR}$.  Our assumptions for $T_\mathrm{d}$ and $\beta$ agree with the results of \citet{Beelen06} for three $350\,\mathrm{\mu m}$-observed QSOs between redshifts of 1.8--2.6 with a mean fitted $T_\mathrm{d}\simeq(35\pm7)$\,K when $\beta$ is fixed to 1.6.  
Note that assuming $T_\mathrm{d}=50$\,K effectively increases $L_\mathrm{FIR}$ by a factor 
of $\simeq$\,2.5 \citep{WangWagg07}.
We have corrected the observed 1.2\,mm flux of J164914.9 by $\simeq$--1.5\,mJy, 
assuming a radio spectral index of 0.1 (see \S \ref{qd25}) for synchrotron emission, before computing $L_\mathrm{FIR}$.  

These $L_\mathrm{FIR}$ estimates are comparable to those of SMGs \citep{Greve05} which is not surprising given the similar submm flux selection criterion.  These assume that $L_\mathrm{FIR}$ is produced predominantly by star-formation as opposed to AGN activity (see discussion in \citealt{Omont03}).  Recent observations have shown that the $L_\mathrm{FIR}$ in QSOs is not generally contaminated by AGN, but is due to star formation, even in the most powerful QSOs (e.g.\ \citealt{Lutz07}; \citealt{Wang07}), although there are clear exceptions (e.g.~\citealt{Weiss07}).

This calculation yields a median of $L_\mathrm{FIR}=(8.0\pm1.9)\times10^{12}$\,L$_\odot$ for the submm-detected QSOs.  
We have converted $L_\mathrm{FIR}$ into a star-formation rate (SFR) for our sample following \citet{Kennicutt98}; 
SFR=$1360\pm320$\,M$_\odot\,\mathrm{yr}^{-1}$.  This relation assumes a \citet{Salpeter} initial mass function 
(IMF) and applies to starbursts with ages less than 100\,Myr.  See \citet{Omont01} for a discussion of 
the full range of conversion factors taking into account the uncertainty in the burst age, IMF, and metallicity.

The star formation efficiency (SFE) is a measure of how effective a galaxy is at converting its gas into stars and can be represented by the continuum-to-line ratio, or $L_\mathrm{FIR}/L'_\mathrm{CO}$, as this ratio presumably traces the star formation rate per total amount of gas in a galaxy.  We find a median SFE=$250\pm100$\,L$_\odot\,\mathrm{(K\,km\,s^{-1}\,pc^{2})^{-1}}$ for the full sample of 
submm-detected QSOs.  

\subsection{SMBH masses}\label{smbhmasses}

We determine $M_\mathrm{BH}$ for our QSOs using the \citet{Greene05} virial $M_\mathrm{BH}$
estimator which calculates the BH mass from the H$\alpha$ or H$\beta$ emission line widths and fluxes.  We can measure these lines from our near-infrared spectra (Fig.~\ref{fig:nirspec}):  the observed emission line properties are 
given in Table~\ref{tab:nir} and the BH masses are given in Table~\ref{tab:derived}.  We find a median $M_\mathrm{BH}$ of $(1.8\pm1.3)\times10^{9}$\,M$_\odot$.  Note that six out of nine of the submm-detected QSOs have 
very large BH masses ($M_\mathrm{BH}\gsim10^{9}$\,M$_\odot$) and are among the largest masses known for local BHs (e.g.~\citealt{Valtonen07}; \citealt{Macchetto97}; \citealt{Humphrey08}), and that two objects in the sample are at the upper envelope of known BHs at any redshift, confirming that these BHs are probably hosted by the progenitors of the most massive present-day ellipticals.

Note that we have not corrected the H$\alpha$ fluxes for extinction, since extinction corrections are rarely applied to broad-line fluxes when estimating virial BH masses, even when rest-frame UV lines are used.   Additionally, \citet{Alexander07} do not find evidence for large amounts of extinction of the broad-line region in their sample of broad-line SMGs (the mean extinction is only $A_V\sim1.2$\,mags, assuming an intrinsic ratio of H$\alpha$/H$\beta$=3.1 and a \citet{Calzetti00} reddening law). Applying this mean extinction correction to our sample would increase the BH masses by $\sim15$\% (0.1\,dex).  We also investigate whether the Eddington ratios, $\eta$, for our submm-detected QSOs are similar to more typical optically selected QSOs in the same redshift range.  Following \citet{McLure04}, we calculate the mass accretion rates and Eddington ratios, $\eta$, for our submm-detected QSOs and find a range in $\eta$ consistent with typical QSOs in the same redshift range ($\eta\sim0.1$--1; \citealt{McLure04}), with a median $\eta\approx0.7$, which is slightly higher than, but consistent with typical QSOs given the large uncertainties involved.  

\begin{table*}
\begin{minipage}{1.0\textwidth}
\scriptsize
\caption{Physical properties of our submm-detected QSO sample derived from the near-infrared and CO observations.}
\label{tab:derived}
\hspace{-0.2in}
\begin{tabular}{lcccccc}
\hline
\multicolumn{1}{l}{Source} & \multicolumn{1}{c}{$L'_\mathrm{CO}$$^{a}$} & \multicolumn{1}{c}{$L_\mathrm{FIR}^{b}$} &  \multicolumn{1}{c}{SFR$^{c}$} & \multicolumn{1}{c}{$M_\mathrm{gas}^{d}$} & \multicolumn{1}{c}{$M_\mathrm{dyn}sin^{2}(i)^{e}$} &  \multicolumn{1}{c}{$\mathrm{log}\,M_\mathrm{BH}$} \\
\multicolumn{1}{c}{} & \multicolumn{1}{c}{($\times10^{10}$\,K\,km\,s$^{-1}$\,pc$^{2}$)} & \multicolumn{1}{c}{($\times10^{13}$\,L$_\odot$)} & \multicolumn{1}{c}{(M$_\odot$\,yr$^{-1}$)} & \multicolumn{1}{c}{($\times10^{10}$\,M$_\odot$)} & \multicolumn{1}{c}{($\times10^{10}$\,M$_\odot$)} & \multicolumn{1}{c}{(M$_\odot$)} \\
\hline
HS1002+4400           & $4.2\pm0.8$ & $1.1^{+1.3}_{-0.9}$ & $1900^{+2200}_{-1600}$ & $3.4\pm0.7$ & $3.3^{+1.9}_{-1.5}$ & $10.14\pm0.2$ \\
RXJ121803.82+470854.6 & $\leq2.4$ & $0.6^{+0.8}_{-0.6}$ & $1100^{+1400}_{-960}$ & $\leq1.9$ & -- & -- \\
SMM\,J123716.01+620323.3 & $\leq0.6$ & $0.5^{+0.7}_{-0.4}$ & $930^{+1200}_{-620}$ & $\leq0.5$  & -- & $8.1\pm0.5$ \\
RXJ124913.86--055906.2 & $3.6\pm1.0$ & $0.7^{+0.9}_{-0.6}$ & $1200^{+1500}_{-1000}$ & $2.9\pm0.8$ & $9.7^{+7.0}_{-5.1}$ & $9.76\pm0.2$ \\
SMM\,J131222.35+423814.1 & $1.2\pm0.4$ & $0.3^{+0.4}_{-0.2}$ & $520^{+670}_{-360}$ & $1.0\pm0.3$ & $2.5^{+2.4}_{-1.6}$ & $8.2\pm0.5$ \\
VV96 J140955.5+562827 & $8.2\pm0.6$ & $2.7^{+2.9}_{-2.6}$ & $4600^{+5000}_{-4500}$ & $6.6\pm0.5$ & $0.8^{+0.2}_{-0.1}$ & $9.28\pm0.2$ \\
VV96 J154359.3+535903 & $3.1\pm0.7$ & $1.0^{+1.3}_{-0.7}$ & $1700^{+2200}_{-1200}$ & $2.5\pm0.6$ & $2.0^{+1.2}_{-0.9}$ & $10.13\pm0.15$ \\
HS1611+4719           & $5.1\pm0.8$ & $1.2^{+1.4}_{-1.0}$ & $2100^{+2400}_{-1400}$ & $4.1\pm0.6$ & $0.4^{+0.2}_{-0.1}$ & $9.26\pm0.2$ \\
MM\,J163655+4059        & $\leq0.8$ & $0.6^{+0.7}_{-0.4}$ & $950^{+1200}_{-690}$ & $\leq0.6$ & -- & $8.4\pm0.5$ \\
VV96 J164914.9+530316 & $\leq2.2$ & $0.8^{+0.9\,f}_{-0.7}$ & $1400^{+1600}_{-1100}$ & $\leq1.8$ & -- & $9.16\pm0.15$ \\
\hline
Notes: \\
\end{tabular}

(a) This quantity was obtained by fitting a Gaussian distribution to the CO spectrum (see text)\\
(b) Derived assuming $\beta=1.5$ and $T=40$\,K (see text) \\
(c) Derived following \citet{Kennicutt98} (see text) \\
(d) Derived assuming $\alpha=0.8$ \\
(e) Derived assuming a disk radius of 2\,kpc\\
(f) 1.2\,mm flux is corrected for synchrotron emission before computing $L_\mathrm{FIR}$\\

\end{minipage}
\end{table*}
\normalsize

\section{Comparison of the submm-detected QSOs and SMGs}\label{discuss}

We now compare the CO properties of the submm-detected QSOs with SMGs.
In order to fairly compare the SMG and QSO samples,
we have discarded three strongly lensed galaxies
from the CO-observed SMG sample from \citet{Greve05} that would not have been detected with $\gsim5$\,mJy at 850\,$\mu$m.
These would lie below our sample flux limit except for the boost from the
lensing magnification, and they thus probe an intrinsically fainter
population.  This leaves a sample of 17 SMGs (of which 11 are CO-detected) from 
\citet{Greve05} with a median observed 850$\,\mu$m flux of 8.2\,mJy.

We compare median values of the gas masses, CO linewidths, SFEs, M$_\mathrm{dyn}$, and M$_\mathrm{BH}$/M$_\mathrm{sph}$.  
The error bars on these median values represent the spread of values, and have been obtained by taking the standard deviation of the median values from 50 trials of randomly drawing out the appropriate number of values out of each population with replacement.  We note that the dominant uncertainties in $M_\mathrm{gas}$, $L_\mathrm{FIR}$, and $M_\mathrm{dyn}$ are likely the assumed CO-to-gas conversion factor, the assumed $\beta$ and $T_\mathrm{d}$ in the SED fitting, and the assumed CO radius and inclination angle, respectively, and have not been included in each error budget, although the relative $M_\mathrm{gas}$, $L_\mathrm{FIR}$, and $M_\mathrm{dyn}$ in our samples will be correct if the same assumptions hold for all our SMGs and submm-detected QSOs.

\subsection{The gas masses and star-formation efficiencies of submm-detected QSOs and SMGs}

\begin{figure}
\epsfig{file=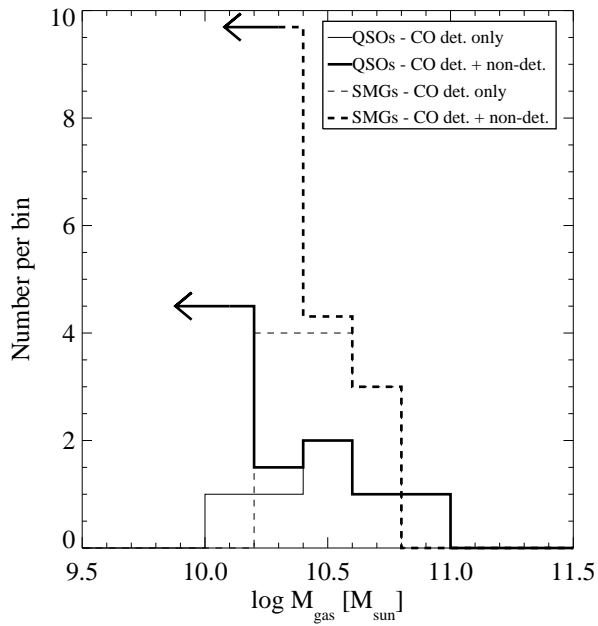,width=0.5\textwidth}
\caption{Distribution of the gas masses of the submm-bright QSO and SMG populations.  We show an estimate of the cold gas mass distribution of submm-bright QSOs and SMGs, including the CO detections {\it and} the gas mass upper limits from the non-detections, constructed using the Avni estimator for censored data (see text).  Note that the bins containing points log\,$M_\mathrm{gas}<10.2$ and 10.4 for QSOs and SMGs, respectively, have been made arbitrarily large here since the Avni estimator requires $\geq1$ detection in the lowest bin.  The Gehan statistic reveals that the distributions are not significantly different.}
\label{fig:gas}
\end{figure}

The median gas mass of our CO-detected QSOs is $M_\mathrm{gas}=(3.4\pm0.8)\times 10^{10}$\,M$_{\odot}$ (see Fig.~\ref{fig:gas}).  This is similar to the median gas mass of the comparison sample of 11 CO-detected SMGs (M$_\mathrm{gas}=(3.0\pm0.5)\times 10^{10}$\,M$_{\odot}$).  Including the upper limits from the CO-undetected objects in each sample yields median gas masses of $(2.5\pm0.7)\times 10^{10}$ and $(2.4\pm0.5)\times 10^{10}$\,M$_{\odot}$ for submm-detected QSOs and SMGs, respectively. Not surprisingly, we find that the gas mass distributions of both samples are statistically indistinguishable at the present sample sizes, as we show below.

Given the moderate fraction of non-detections in both samples, we employ survival analysis (see e.g.\ \citealt{Feigelson85}) to estimate the intrinsic distributions of the gas masses of submm-bright SMGs and QSOs.  This includes the detections \textit{and} upper limits, assuming that the data are censored at random (i.e.~that the chance of only a gas upper limit being available for an object is independent of the true value of the gas mass).  Given the current detection limit and small dynamic range in our sample, there is no strong apparent trend with the selection criteria (i.e.~with submm flux) versus whether a particular source is detected or not, and thus the assumption of the data being randomly censored seems reasonable.  We employ the widely-used product-limit Avni estimator (\citealt{Avni80}; see also \citealt{Wall}) to construct a maximum-likelihood-type reconstruction of the true distribution of gas masses in the QSO and SMG samples (see Fig.~\ref{fig:gas}).  We now test whether or not our censored samples of SMG and QSO gas masses are likely to have been drawn from the same distribution using the non-parametric Gehan test, following \citet{Feigelson85}.  The Gehan statistic yields $L_\mathrm{n}=42\pm39$ (equivalent to $\simeq1\,\sigma$), revealing that the difference between the gas mass distributions for SMGs and QSOs is about 35\% likely due to chance.  We consider 35\% to be a high value, and therefore conclude that the gas masses of our QSO sample are indistinguishable from that of the SMG sample at the same epoch of $z\sim2$--3 for the high mass end (where we are the most sensitive).

\begin{figure}
\epsfig{file=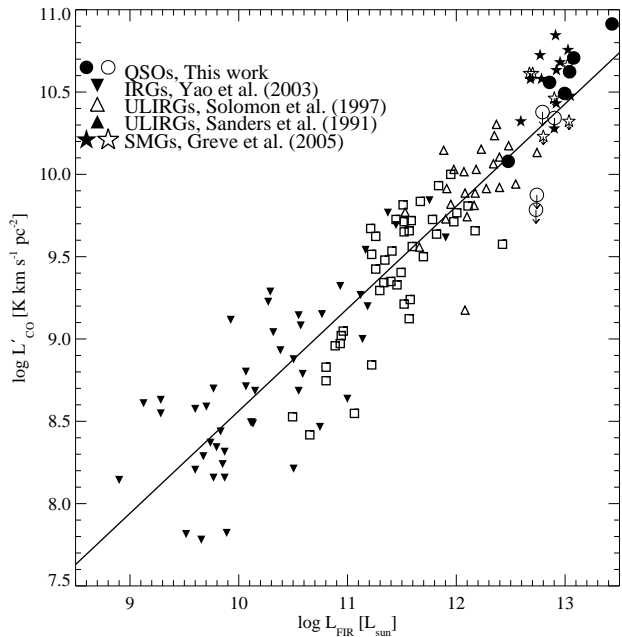,width=0.5\textwidth}
\caption{Comparison of the CO and far-infrared luminosities for QSOs, LIRGs, ULIRGs, and SMGs.  The line is the best-fitting relation with a form of log\,$L'_\mathrm{CO}=\alpha\,\mathrm{log}\,L_\mathrm{FIR}+\beta$ to the LIRGs, ULIRGs and SMGs from \citet{Greve05}.  The CO-observed SMGs and QSOs appear to occupy the same part of the diagram (filled circles/stars represent CO-detections and open circles/stars are CO-nondetections), and lie just above the relation though the offset is not significant given the large errors in $L_\mathrm{FIR}$.  This is a particularly useful diagnostic, since it does not depend on the CO-to-gas conversion factor.}
\label{fig:sfe}
\end{figure}

Within our QSO sample we find a broad similarity in the star formation efficiency of the CO-detected and undetected QSOs (see \S~\ref{lfircalc}), assuming conservatively that the CO-undetected subset lies just below the detection threshold.
A Kolmogorov-Smirnov (KS) test reveals a $25\%$ chance that the two subsamples are drawn from the same parent population, indicating that we have not detected a statistically significant difference in the SFE between the CO-detected and undetected QSOs in our small sample given the current detection limits.  What is the cause then of a given submm-detected QSO in our sample, which all have similar $L_\mathrm{FIR}$, being either CO-detected versus CO-undetected?  A possible explanation could be that the CO-detected submm-detected QSOs are physically larger and contain proportionally more massive CO reservoirs than the CO-undetected QSOs, although time-dependent star formation, differences in the SEDs or filling factors could also be causes for the scatter in the $L_\mathrm{FIR}$--$L'_\mathrm{CO}$ correlation (see Fig.~\ref{fig:sfe}).

Locally, L$'_{\rm CO}$ increases with L$_{\rm FIR}$ for (U)LIRGs, with the 
\citet{Greve05} sample of SMGs extending this 
trend out to the highest far-infrared luminosities 
($\gtrsim10^{13}$L$_{\odot}$).  
For comparison, in Fig.~\ref{fig:sfe} we have plotted our QSOs on the 
L$'_{\rm CO}$--L$_{\rm FIR}$ diagram along with LIRGs, ULIRGs and SMGs.  
The CO-observed SMGs and QSOs seem to lie along the relation within the considerable 
uncertainties in their far-infrared luminosities, suggesting
 that the SMGs and QSOs are undergoing similar star formation modes, despite the QSOs hosting luminous AGN activity.

To quantify this we compare the SFE of the QSOs to the SMGs, where $L_\mathrm{FIR}$ for the SMGs has been calculated using the method outlined in \S \ref{lfircalc} and their $L'_\mathrm{CO}$ are taken from \citet{Greve05}.  As Fig.~\ref{fig:sfe} shows, the SMGs appear to have similar SFE with respect to QSOs on average, although this probably reflects the similar submm flux selection criteria.  Given that the median $L_\mathrm{FIR}$ and $M_\mathrm{gas}$ of the QSO and SMG samples are similar, $(8.0\pm1.9)\times 10^{12}$\,L$_{\odot}$ and $(7.9\pm0.8)\times 10^{12}$\,L$_{\odot}$ and $(2.5\pm0.7)\times 10^{10}$\,M$_{\odot}$ and $(2.4\pm0.5)\times 10^{10}$\,M$_{\odot}$ (including all CO-detected and -undetected QSOs and SMGs), respectively, the SFEs are also similar: medians of $250\pm100$ and $260\pm50$\,L$_\odot\,\mathrm{(K\,km\,s^{-1}\,pc^{2})^{-1}}$, for QSOs and SMGs respectively.  We perform a Gehan test on these censored data sets and find $L_\mathrm{n}=56\pm35$ ($\simeq1.6\,\sigma$), revealing that the difference between the distribution of SFE in SMGs and QSOs is about 15\% likely due to chance, indicating that we have not detected a significant difference in the SFE between our small samples of SMGs and QSOs.

\subsection{The CO line profiles and dynamical masses of submm-detected QSOs and SMGs}\label{fwhm_discuss}

A comparison between the observed CO line profiles of SMGs and QSOs requires care as we know that there is an 
inherent difference in the types of lines seen in SMGs versus QSOs: five of the 11 SMGs detected in CO in our 
comparison sample show hints of double-peaked profiles (\citealt{Greve05}; \citealt{Tacconi06}), while none of our QSO 
CO emission lines appear double-peaked (which is perhaps not surprising given our small sample size and low SNR data).  
These double-peaked profiles could indicate merger activity (as likely for SMGs from high-resolution observations), 
or a rotating gas disk/ring (e.g.\ \citealt{Genzel03}).  In both cases, measuring the separation between the peaks is a more 
appropriate measure of the dynamics of the merger or rotating disk system than the FWHM of the individual peaks.

In cases where double-peaked profiles are present we need to fit two Gaussian profiles 
simultaneously in order to derive the velocity offsets of the peaks.  
We have therefore refit all of the SMGs and QSOs with single and double Gaussian profiles using 
the $\chi^{2}$ statistic as a measure of the goodness-of-fit.  We find that 4/11 of the 
CO-detected SMGs are better fit at the $>3.5\,\sigma$ level with a double Gaussian profile over a single Gaussian profile \citep{Greve05}.  
None of the submm-detected QSOs are significantly better fit by double Gaussian profiles, although we note that the random chance of drawing six single peaked profiles (with replacement) out of a sample containing seven single-peaked and four double-peaked profiles (i.e.~the SMG sample above) is 7 percent.  The fact that the QSO profiles appear to be single-peaked could indicate that our data are too low SNR to detect double-peaked profiles (especially for a merger, where both the width and peak strength of the lines can be different), or that we are viewing the QSOs almost face-on\footnote{An intrinsically double peaked profile could appear single-peaked if the velocity offset between the peaks is small enough (e.g.~at high SNR a close merger could show an asymmetric line profile, hinting at two components).  We attempt to put a limit on the inclination angles in our data by investigating a simple test case.  We take two Gaussian distributions with $\sigma=150$--$220\,\mathrm{km\,s^{-1}}$ (similar to those found for double-peaked SMGs) and equal line fluxes and shift them to a velocity separation such that the FWHM of a single-Gaussian profile fit is $\simeq550\,\mathrm{km\,s^{-1}}$ (our QSO sample median FWHM), yielding typical velocity separations of 300--$150\,\mathrm{km\,s^{-1}}$. This translates into a conservative limit on the inclination of the system of $i\lesssim40^{\circ}$ by using the relation $v_\mathrm{obs}=v_\mathrm{circ}\,sin(i)$ and taking $v_\mathrm{obs}\simeq250\,\mathrm{km\,s^{-1}}$ and $v_\mathrm{circ}=400\,\mathrm{km\,s^{-1}}$ (the latter value is what is observed typically for SMGs by fitting double Gaussian profiles).} if the CO is contained in a disk, or that they are at a late stage of a merger, or that they are not in the course of a merger.

\begin{figure}
\epsfig{file=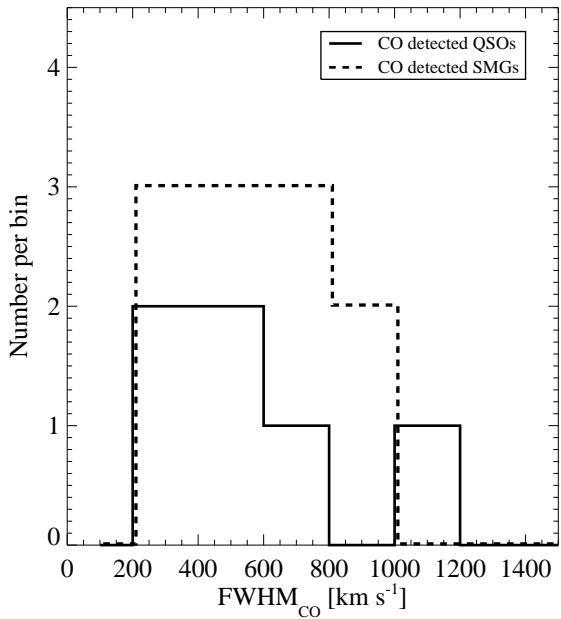,width=0.5\textwidth}
\caption{Comparison of the distributions of CO FWHM for 11 SMGs and 6 QSO with detected CO emission.  The distributions have medians of $(530\pm110)\,\mathrm{km\,s^{-1}}$ (for a mix of single and double Gaussian profile fits), and $(550\pm180)\,\mathrm{km\,s^{-1}}$ for SMGs and QSOs, respectively.  A KS test reveals a 95\% probability that the SMG and QSO CO FWHM distributions are drawn from the same parent population (see text).  The histograms have been offset slightly for clarity.}
\label{fig:fwhm}
\end{figure}

For our six submm-detected QSOs we compare the CO FWHMs and dynamical masses 
with the CO-detected sub-sample of 11 SMGs taken from the literature (\citealt{Greve05}; \citealt{Tacconi06}). 
The median FWHM of the six CO-detected QSOs and seven single-Gaussian profile fit SMGs are $(550\pm180)$ and $(600\pm130)\,\mathrm{km\,s^{-1}}$, respectively.  When we include the double Gaussian profile fits in the median estimate, taking the separation between the peaks as a proxy for the dynamics of the system, 
the median for the SMGs slightly decreases to $(530\pm110)\,\mathrm{km\,s^{-1}}$.  We have plotted the histogram of the submm-detected QSO and SMG line widths for comparison in Fig.~\ref{fig:fwhm} (c.f.~fig.~2 in \citealt{CarilliWang06}).

The CO FWHM of QSOs and SMGs appear to be similar.  To quantify this 
we use a standard two-sided KS test which reveals a 95\% probability that the two distributions are drawn from the same parent population (this drops to 77\% if only the single-line Gaussian profile SMGs are used).   This is in stark contrast with \citet{CarilliWang06}, who find only a 0.9\% probability for an inhomogeneous sample 
of 12 SMGs and 15 QSOs from a literature compilation (\citealt{Solomon05}).
However, we note that they have used the SMG single-Gaussian profile fit line widths even in cases where clearly double-peaked CO profiles are apparent, leading to artificially broad FWHM in these cases.  Their analysis is also based on a mixed sample over a wide redshift range and includes strongly lensed sources.  Also, the QSOs studied here have larger linewidths than have been reported for other QSOs and they are also those with the lowest SNRs and thus the most difficult to fit.  Therefore the differences in our results for $z\sim2$ QSOs and those in \citet{CarilliWang06} for higher redshift QSOs can probably mainly be attributed to the small sample sizes involved and the somewhat different approaches (rather than a real statistically significant effect) when all possible sources of error are taken into account.

Since the CO line widths of SMGs and QSOs are similar, it may follow that the dynamical masses of SMGs and QSOs are also similar. The uncertainties in this are if the gas is distributed similarly in both populations (i.e.~either a disk or merger scenario), if their sizes ($R$) are the same, and if their average inclination angles ($i$) are similar.  $R\simeq2$\,kpc is believed to be appropriate for the gas distribution in SMGs (e.g.\ \citealt{Greve05}; \citealt{Tacconi06, Tacconi08}).  One of our QSOs, J140955.5, has been observed at PdBI by \citet{Beelen04} in CO(7--6) with a $1.0''\times0.5''$ resolution and was unresolved, giving an upper limit to the gas reservoir of 5\,kpc (though this could be too high a J transition to trace the bulk of the gas reservoir).  No sizes have been measured for our sample of QSOs, although the CO emission seems to be confined to a compact region in other samples of high-z QSO \citep{Riechers06} as inferred by comparing a range of CO line ratios (see also \citealt{Walter04} who observe gas distributed over a 2.5\,kpc radius and \citealt{Riechers08}).  This agrees with the recent results from \citet{Maiolino07} who find that CO emission in a $z\simeq5$ QSO is unresolved with a beam size of $\sim1$\,arcsec, implying that the molecular gas is contained within a compact region with a radius of $<3$\,kpc.  The QSO from \citet{Maiolino07} was observed in a high J transition, CO(5--4), and one might worry that such high transitions might not trace the bulk of the reservoir.  CO(1--0) and (2--1) observations of other $z\gsim4$ QSOs (e.g.~\citealt{Riechers06}) have confirmed that the high J transitions do trace the bulk of the reservoir, and thus provide good size constraints (this is in contrast to what is observed in most nearby galaxies).  Spatially resolved observations exist for local QSOs, albeit the data are sparse:  \citet{DownesSolomon98} measure a $R\gtrsim1$\,kpc disk in Mrk 231; \citet{Staguhn04} find a $R\gtrsim1$\,kpc ring-like molecular gas distribution; and \citet{Krips07} estimate $R\simeq3\pm1$\,kpc for the CO emission in a low luminosity QSO.  Therefore, assuming 2--3\,kpc radius regions will probably still be reasonable once these galaxies are resolved.

While the average orientation of the SMGs is likely to be close to the value expected for random orientations, $i\simeq30^{\circ}$, since they are not prone to orientation selection effects (see \citealt{CarilliWang06}), the submm-detected QSOs could be orientated preferentially with respect to the sky plane due to their selection as optically luminous QSOs in the first place.  We find evidence for this from a comparison of the [O{\sc iii}] line width (see Table~\ref{tab:nir}), a proxy for the measure of the stellar velocity dispersion (supporting the idea that the narrow-line region gas is in orbital motion in the gravitational potential well of the bulge; \citealt{Nelson96}; \citealt{Nelson00}; \citealt{Bonning05}), with the CO line width.  We find that the [O{\sc iii}] FWHM are a median factor of $1.9\pm0.5$ broader than the CO line profiles (with a large scatter; ratios range from 0.6--4.3), suggesting indeed that inclination effects could be important in the CO line widths of QSOs (see also \citealt{Shields06}).  The lack of the double-peaked CO profiles in the submm-detected QSOs could be interpreted as a later phase of a merger in the typical proposed evolutionary sequence, with SMGs representing an earlier phase of the merger.  We also know that at least two of our QSOs must have $i\lesssim20$ (alternatively the assumed standard CO-to-H$_2$ conversion factor could be too high or else the assumed 2\,kpc radius is too small) or else the cold gas mass estimates exceed the dynamical masses (even given the large error bars; see Table~\ref{tab:derived}).  Conversely, some of our broadest CO line detections suggest that their inclination angles might be larger than $20^{\circ}$.  Other constraints come from the obscured:unobscured AGN ratio ($\simeq5$--10) for a sample of SMGs, where \citet{Alexander07} suggest that QSOs could be seen close to the line of sight ($i=18$--$25^{\circ}$).  We thus adopt the best available estimates of the average inclination angles for SMGs and QSOs of $30^{\circ}$ and $20^{\circ}$, respectively, in calculating the dynamical masses below. High-resolution CO observations could help to place tighter constraints on the inclination angles of our systems. 

Assuming a 2\,kpc radius and an average inclination angle of $20^{\circ}$ for the submm-detected QSOs and $30^{\circ}$ for the SMGs, 
we find median dynamical masses, M$_\mathrm{dyn}\propto \mathrm{FWHM_{CO}}^{2}\,R\,csc^{2}(i)$, of M$(<2$\,kpc$)\simeq(2.1\pm1.4)\times 10^{11}$ and $(0.9\pm0.4)\times 10^{11}$\,M$_\odot$, respectively.
This results in gas-to-dynamical mass fractions of $\sim15$\% and 30\% for QSO and SMG populations, respectively.  The gas mass accounts for a significant fraction of the dynamical mass of both classes, indicating that the host galaxies are at an early evolutionary stage.

\begin{figure}
\epsfig{file=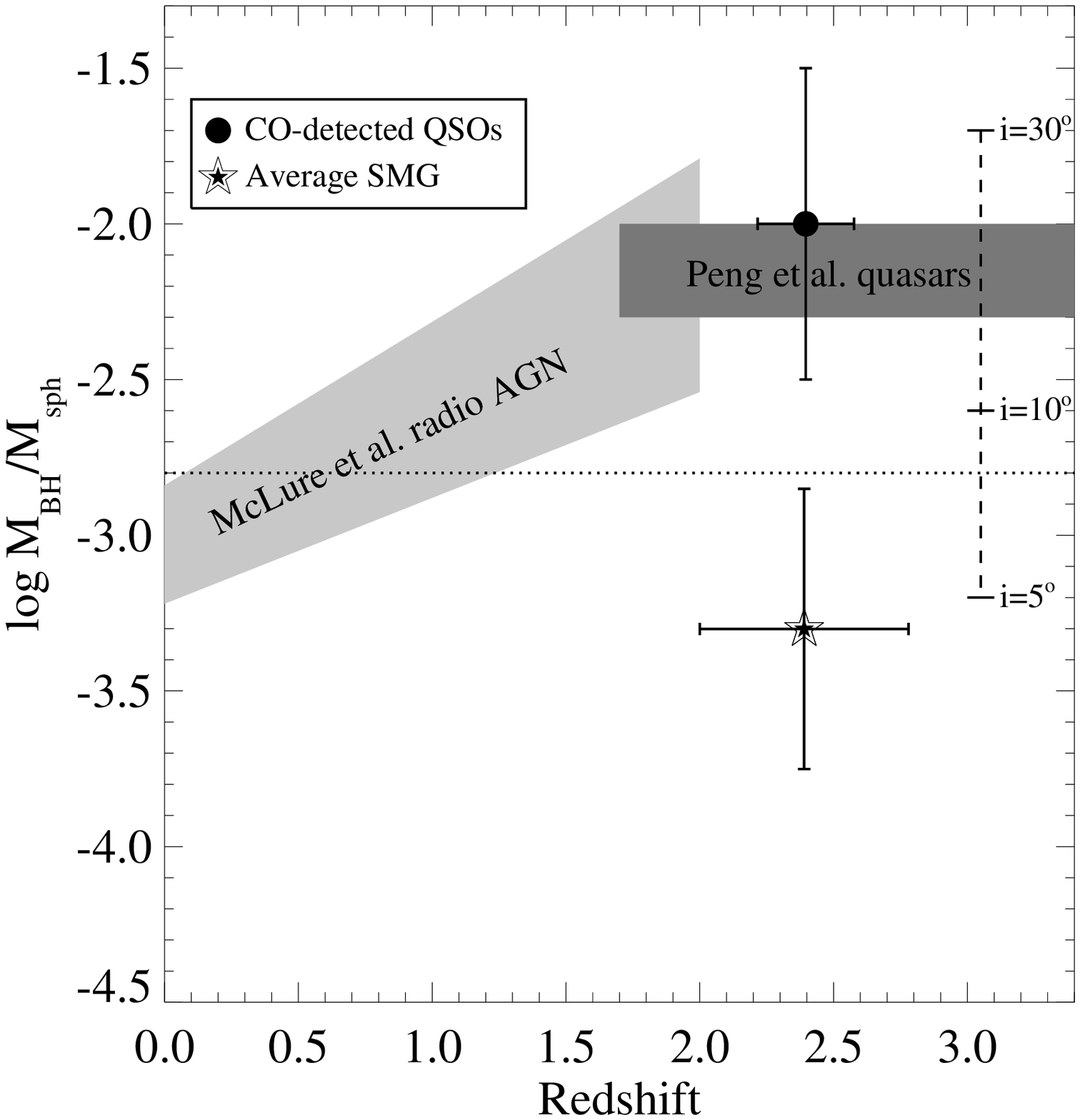,width=0.5\textwidth}

\caption{The variation in $M_\mathrm{BH}/M_\mathrm{sph}$ with redshift,
  $z$, for different populations of galaxies and AGN.  We show the
  ratio measured for BHs in local spheroids (dotted line; \citealt{Haring04}) and those inferred for 
high-redshift radio-loud AGN \citep{McLure06}, QSOs
\citep{Peng06}, and SMGs \citep{Alexander07}.  We compare these to the ratio
determined for our CO- and submm-detected QSOs.  We find good agreement
between our estimate and that of \citet{Peng06} for similar, optically
luminous QSOs at $z\sim 2$, supporting the assumptions used in our
analysis, although we show how our estimate changes for different assumed inclination angles (dashed lines).  
Similarly, the typical QSO detected in our survey has a
significantly higher $M_\mathrm{BH}/M_\mathrm{sph}$ ratio than
inferred for $z\sim 2$ SMGs.}
\label{fig:ratio}
\end{figure}

\subsection{The BH-spheroid ratios of submm-detected QSOs}\label{bulge}

We place a constraint on the spheroid mass by taking the dynamical mass
estimate from \S \ref{fwhm_discuss}, assuming $i=20^{\circ}$ and a spheroid
radius of 2\,kpc or less (as we assumed for the CO emission),
yielding a median spheroid mass of $M_\mathrm{sph}=(2.1\pm1.4)\times
10^{11}$\,M$_\odot$. In principle one should subtract the gas and BH masses, although we note that doing so 
makes a negligible difference in general, given the large error bars on $M_\mathrm{dyn}$, and we also have 
an additional complication that two of our CO-detected QSOs have $M_\mathrm{gas}>M_\mathrm{dyn}$.
Other uncertainties to our $M_\mathrm{sph}$ estimates include the possibility that the cold gas in
submm-detected QSOs is more widely distributed (e.g.\ due to entrainment in
outflows) than we have assumed here ($\gsim2$\,kpc), which would mean
that all of our dynamical (and hence spheroidal) masses have been underestimated.  
Similarly, if the spheroid is significantly more extended than the
CO emission then the spheroid masses will be larger.  For example,
assuming a characteristic radius of $\sim 10$\,kpc for the spheroid
would result in an increase in the mass of a factor of $\sim 5$ times.

Based on these assumptions we estimate
$M_\mathrm{BH}/M_\mathrm{sph}\sim (9^{+21}_{-6})\times10^{-3}$
for our QSO sample.
This is approximately an order of magnitude larger than the local
ratio of $(1.4\pm0.4)\times10^{-3}$ for the BHs in local spheroids
(Fig.~\ref{fig:ratio}; \citealt{Marconi03}; \citealt{Haring04}).  
Thus our CO survey of $z\sim 2$ QSOs appears to support the claim that
$M_\mathrm{BH}/M_\mathrm{sph}$ for $z>1$ AGN is higher than
observed locally (e.g.\ \citealt{Peng06};
\citealt{McLure06}).  Our $M_\mathrm{BH}/M_\mathrm{sph}$ ratio could be made to match the local 
relation by either decreasing the inclination angles to $\sim 6^{\circ}$ or by adopting a 
spheroid extent ten times larger than we have assumed.
Our results and those of \citet{Peng06} suggest that
BH growth may occur more rapidly than spheroid formation in QSOs at $z\sim 2$ (although both studies could 
be prone to strong selection effects; see \citealt{Alexander07}). Similar CO-based studies of
rarer individual QSOs at higher redshifts also support
this conclusion (e.g.\ \citealt{Walter04}; \citealt{Maiolino07}).

This result contrasts strongly with the $M_\mathrm{BH}/M_\mathrm{sph}$ 
claimed for SMGs, which lie just below the local ratio (Fig.~\ref{fig:ratio}; \citealt{Alexander07}). 
However, as mentioned in
\S \ref{smbhmasses}, six out of nine of our submm-detected QSOs have
very massive BHs ($M_\mathrm{BH}\gsim10^{9}$\,M$_\odot$) which 
are far more massive than those inferred for typical SMGs (\citealt{Alexander07}; $M_\mathrm{BH}\approx10^{8}$\,M$_\odot$) and would place
them among the largest known masses for local BHs.  They are therefore likely to be considerably rarer than typical SMGs.  
To make a fairer comparison between SMGs and submm-detected QSOs we need to focus on those CO-detected submm-detected QSOs with
$M_\mathrm{BH}$ more similar to that found for SMGs.  We have three submm-detected QSOs 
in our sample with BHs more similar to those of SMGs, unfortunately 
two of these are undetected in CO, so we cannot determine their
dynamical masses.  Nevertheless, the one example with a marginal CO detection,
SMM\,J131222.35, has an implied spheroid mass of
$M_\mathrm{sph}=2.1\times10^{11}$\,M$_\odot$ (assuming $i=20^{\circ}$)
and hence $M_\mathrm{BH}/M_\mathrm{sph}\simeq9\times10^{-4}$. This is
an order of magnitude below the ratio for the more optically luminous QSOs and places
SMM\,J131222.35 approximately on the local relation (Fig.~\ref{fig:ratio}).  While based on
a single source marginally detected in CO, this
result hints that submm-detected QSOs with
$M_\mathrm{BH}\simeq10^{8}$\,M$_\odot$ might be the SMG--QSO
`transition' objects we are interested in.   We discuss this
possibility further in
\S \ref{evolution}.

%
%
%
\section{Discussion of the evolutionary status of submm-detected QSOs}\label{evolution}

In this section we discuss the various observations which constrain the
evolutionary relationship between SMGs and submm-detected QSOs.  For
the purposes of this discussion, we adapt the proposed evolutionary
pathway of \citet{Sanders88} and so test the scenario where the SMG
population evolves through a submm-detected QSO phase, into a
submm-undetected QSO and finally a passive elliptical.  Now we ask --
are the observed properties of the SMGs and submm-detected QSOs
consistent with their observed gas depletion, amount of BH growth,
and space densities required to link them in this evolutionary sequence?

To provide a characteristic timescale for our discussion we first
determine the lifetime of the submm-detected QSO phase. We assume that
the far-infrared luminosities of the submm-detected QSOs arise from
dust-obscured star formation and that the molecular gas reservoirs we
detect via their CO emission are the fuel for this star formation.  In
this way we determine that the QSOs have enough cold gas to sustain
their current episode of star formation (assuming a star formation efficiency of 100\%) for
$\tau_\mathrm{depletion}\sim
\mathrm{M_{gas}/SFR}\sim2.5\times10^{10}$\,M$_{\odot}/1360$\,M$_\odot\,\mathrm{yr}^{-1}\sim20$\,Myr.
$\tau_\mathrm{depletion}$ is on the order of a couple of dynamical
times within our assumed CO radius of 2\,kpc
($\tau_\mathrm{dyn}=R/v_\mathrm{c}<9$\,Myr), which could be defined as
a `maximal starburst' (e.g.\ \citealt{Tacconi06}) as is commonly found
with other populations of starburst galaxies such as ULIRGs at
low-redshift and SMGs and other QSOs at high-$z$.  Since we are most likely
catching the QSOs halfway through the current star formation episode, we infer a
submm-detected QSO phase lifetime of
$2\times\tau_\mathrm{depletion}=40$\,Myr.
The reader should of course note that $\tau_\mathrm{depletion}$ and our
estimated lifetime are both lower limits if $L_\mathrm{FIR}$ includes a
significant contribution from AGN activity, or if the starburst is
discontinuous, or if the IMF is more top heavy than Salpeter's IMF, or
if the gas reservoirs contain significant masses of sub-thermal gas
\citep{Greve06}, or if the CO reservoir is widely distributed
on scales $\gg30$\,kpc (not that there is any evidence that such galaxies exist).  

\begin{figure}
\epsfig{file=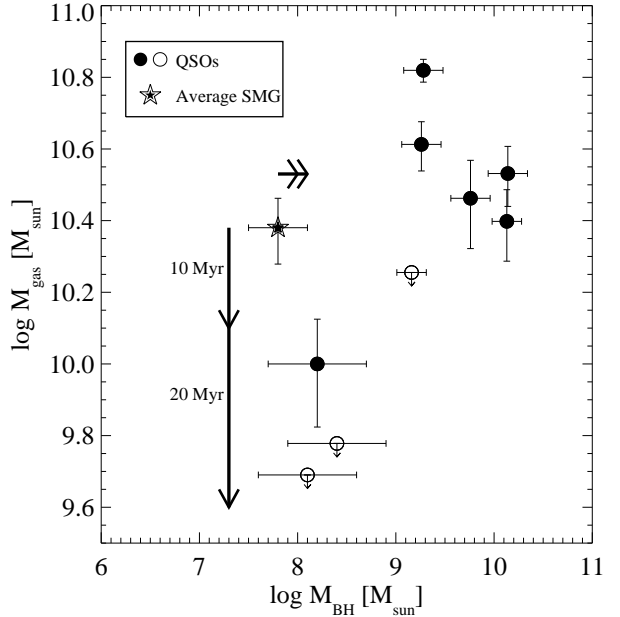,width=0.5\textwidth}
\caption{$M_\mathrm{gas}$ as a function of $M_\mathrm{BH}$ for our
sample of CO-observed submm-detected QSOs.  Recall that the quoted
errors for $M_\mathrm{gas}$ do not include the uncertainty on the
CO-to-gas conversion factor.  The arrows (offset slightly for clarity)
indicate the movement of an average SMG in terms of its gas mass
depletion and synchronous BH growth in arbitrary scalable timecales of
10 and 20\,Myr, assuming a SFR of 1000\,M$_\odot$\,yr$^{-1}$ and
Eddington-limited BH growth.  This demonstrates that an SMG could
evolve into $M_\mathrm{BH}\simeq10^{8}$\,M$_\odot$,
$M_\mathrm{gas}\simeq0.6\times10^{10}$\,M$_\odot$ submm-detected QSOs
in a reasonable timescale, whereas an average SMG would need
substantially ($\sim10\times$) more gas and more time
($\gsim\,\mathrm{SMG\,\,lifetime}$) to evolve into the
$M_\mathrm{BH}\simeq4\times10^{9}$\,M$_\odot$,
$M_\mathrm{gas}\simeq3.0\times10^{10}$\,M$_\odot$ submm-detected QSOs.
This suggests that the submm-detected QSOs hosting BHs of
$M_\mathrm{BH}\simeq10^{8}$\,M$_\odot$ with $M_\mathrm{gas}\lesssim
1\times10^{10}$\,M$_\odot$ comprise `transition objects' that we can
use to probe the intermediary evolutionary stage between the SMG and
luminous QSO phases, while the rarer more luminous QSOs with
$M_\mathrm{BH}\gtrsim 10^{9}$\,M$_\odot$ are not related to typical
SMGs.}
\label{fig:dave}
\end{figure}

Next we note that in our submm-detected QSO sample CO is
typically detected in only the optically luminous QSOs (see
Fig.~\ref{fig:selection}), i.e.\ those QSOs with the most massive BHs
($M_\mathrm{BH}>10^{9}$\,M$_\odot$; see Fig.~\ref{fig:dave}).  As we
discussed in \S \ref{bulge} these BHs are much more massive than found
in typical SMGs -- is it feasible for the BHs to grow sufficiently
between these two phases?  In Fig.~\ref{fig:dave} we plot
$M_\mathrm{gas}$ as a function of $M_\mathrm{BH}$ for our sample of
submm-detected QSOs and for an `average' SMG ($M_\mathrm{BH}$ from
\citealt{Alexander07} and $M_\mathrm{gas}$ from \citealt{Greve05}).  We
indicate gas consumption and BH growth evolutionary tracks for the
average SMG on timescales of 10 and 20\,Myr, assuming a SFR of
1000\,M$_\odot$\,yr$^{-1}$ and Eddington-limited BH growth.
Fig.~\ref{fig:dave} demonstrates that an SMG could evolve into a
submm-detected QSO with $M_\mathrm{BH}\simeq10^{8}$\,M$_\odot$ and
$M_\mathrm{gas}\simeq0.6\times10^{10}$\,M$_\odot$ (the median $M_\mathrm{gas}$ 
of the CO-undetected QSOs) in a reasonable
timescale, whereas an average SMG would need substantially ($\sim10$
times) more gas and more time ($\gsim\,\mathrm{SMG\,\,lifetime}$) to
evolve into the optically luminous submm-detected QSOs in our sample
with $M_\mathrm{BH}\simeq4\times10^{9}$\,M$_\odot$ and
$M_\mathrm{gas}\simeq3.0\times10^{10}$\,M$_\odot$.

The gas consumption and BH growth timescales suggest that the optically
luminous submm-detected QSOs in our sample cannot be the direct
descendents of typical SMGs.  Indeed the CO-detected optically
luminous QSOs in our sample are extremely luminous AGN
($M_\mathrm{B}\approx -28$, \citealt{Omont03}) with
$M_\mathrm{BH}>10^{9}$\,M$_\odot$ and will most likely evolve into
the most massive elliptical galaxies found in local rich clusters. These QSOs are so rare
($\simeq2\,\mathrm{deg}^{-2}$ over all redshifts; \citealt{Fan01}) that
they are unlikely to be associated with \textit{any} known SMGs, since
$S_{850\,\mathrm{\mu m}}>4.5$\,mJy SMGs at $z=1$--3 have much higher
surface densities ($\simeq600\,\mathrm{deg}^{-2}$; \citealt{Coppin06}) or
space densities ($\simeq10^{-5}$\,Mpc$^{-3}$; \citealt{Chapman05}).
Even correcting for the relative timescales of the two phases still leaves 
an order of magnitude difference in the space densities.  
Based on the relative number or space densities, these optically luminous 
submm-detected QSOs are unlikely to be involved in the SMG/QSO
evolutionary sequence we are interested in testing.


In contrast, the less luminous submm-detected QSOs appear to be consistent with the \citet{Sanders88} 
proposed evolutionary sequence, linking them to typical SMGs (see Fig.~\ref{fig:dave}).
Thus we concentrate on the three submm-detected QSOs (including one marginal CO detection and two upper limits:
SMM\,J131222.35, SMM\,J123716.01 and MM\,J163655)
with BHs more similar to that of SMGs of
$M_\mathrm{BH}\simeq10^{8}$\,M$_\odot$.  The median gas
mass and SFR of this subsample are $0.6\times10^{10}$\,M$_{\odot}$ and
850\,M$_\odot\,\mathrm{yr}^{-1}$, respectively.  Hence, these systems can
sustain their current star formation episode for
$\tau_\mathrm{depletion}\lsim7$\,Myr or a 
total lifetime of the submm-detected QSO phase of $\lsim 15$\,Myrs.
This estimate of the lifetime of submm-detected QSOs can then be used
to relate them to SMGs as follows.  
The incidence of blank-field submm
sources identified with QSOs is $\sim 4$\% \citep{Chapman05}.  In
our test scenario 
this is the fraction of an SMG lifetime in which it appears as an
submm-detected QSO.  Conversely, about 15--25\% of optically
luminous QSOs are detected in the submm \citep{Omont03}.  Assuming that
all bright SMGs go through a QSO phase, and that all
QSOs are formed via SMGs and there is no strong mass or
luminosity dependence of their lifetimes, then
these relative fractions imply that the total QSO
lifetime (i.e.~submm-detected plus submm-undetected) is about $20\%/4\%=5$\,times shorter than the 
average lifetime of an SMG.  Taking our estimated lifetime
of less luminous submm-detected QSOs, $\lsim 15$\,Myrs, these detection
fractions imply total lifetimes for QSOs of $\lsim 70$\,Myrs
and for SMGs $\lsim 350$\,Myrs.
These estimates are consistent with independent estimates of the
lifetimes for QSOs of
$\sim20$--40\,Myr (\citealt{Martini01}; \citealt{Goncalves07})
and $\sim100$--300\,Myr for SMGs from modelling \citep{Swinbank06}. 

There are two other observational links between these QSOs
with $M_\mathrm{B}\approx -25$ ($M_\mathrm{BH}\sim 5\times
10^{8}$\,M$_\odot$) and SMGs.  Firstly, the QSOs 
have space densities at $z\sim 2$ some 5--10 times lower than SMGs (\citealt{Boyle00};
\citealt{Chapman03}).  This ratio of volume densities is  
roughly consistent with the ratio of the relative SMG and QSO lifetime,
$\sim100$--300\,Myrs and $\sim 20$--40\,Myrs,  of $\sim 5$ times.
Moreover, in \S
\ref{bulge} we found a hint that this subset of our submm-detected QSOs lies on
the present-day $M_\mathrm{BH}/M_\mathrm{sph}$ relation and similar to SMGs,
which is consistent with them being `transition objects' between SMGs and QSOs
in the proposed evolutionary sequence in which they would eventually evolve 
into passively evolving spheroids.

Although this conclusion is based on a small sample of QSOs, if the dynamical
and gas properties of these three QSOs are representative of this
subset of submm-detected QSOs, then our results are consistent with 
$\sim L^{\star}$, $M_\mathrm{BH}\simeq10^{8}$\,M$_\odot$ submm-detected QSOs
being `transition objects' in the proposed evolutionary sequence, 
representing a very brief prodigious star
formation phase where the BH is undergoing rapid growth synchronously
with the stellar mass.  Submm-detected QSOs do
not possess sufficiently large gas reservoirs 
(containing less gas than SMGs on average) to sustain the implied
SFRs for very long, providing a possible explanation for why some of
our submm-detected QSOs are undetected in CO (a single weak CO
detection and two non-detections of $M_\mathrm{BH}\simeq10^{8}$\,M$_\odot$ submm-detected QSOs).

Where do the optically luminous submm-detected QSOs fit into this
evolutionary scheme?  
The average SMG would need $\approx1$\,Gyr (more than its
lifetime) in order to grow their BHs by the required factor of
$\gsim60$ to host a $M_\mathrm{BH}\simeq
4\times10^{9}$\,M$_\odot$ BH seen in the optically luminous subset
of submm-detected QSOs.  In addition, these SMGs would need to begin
with gas reservoirs containing $M_\mathrm{gas}\gsim
4\times10^{11}$\,M$_\odot$ in order to leave 
luminous QSOs with detectable reservoirs of 
residual gas.  Alternatively, these rare optically-luminous QSOs could
be periodically `repeating' SMGs, with SMG progenitors with
$M_\mathrm{gas}\gsim10^{11}\,M_\odot$ and a surface density of
$\simeq10\,\mathrm{deg}^{-2}$, which are consuming their gas and
growing their BHs synchronously, but which accumulate additional BH
mass and gas through merging with other SMGs in dense environments
(i.e.\ damp `dry merging'; e.g.\ \citealt{Bell04}).  However, we currently have
not observed any SMGs with such large gas masses, and given their
rarity, we will require large scale submm blank-field surveys such as
those planned with SCUBA-2 to find these potential progenitor SMGs.

Hence our current results split the submm-detected QSOs in
our survey into two categories
(although a bigger sample would probably give a continuous trend):

\smallskip (1) Our observations of the optically luminous
submm-detected QSOs indicate they have
$M_\mathrm{BH}>10^{9}\,M_\odot$.  It is very difficult to
grow the BHs in typical SMGs to such large masses in
the typical lifetimes of SMGs or submm-detected QSOs.
We therefore suggest that these QSOs cannot be related to
the evolutionary cycle of typical SMGs.  Instead, these
intrinsically rare QSOs most likely evolve from an equally
rare subset of the SMG population -- whose surface density is
low enough that it has yet to be detected in current
small-area submm surveys.  Simply based upon the masses of
their BHs, it is clear that these systems can only evolve into
the most massive elliptical galaxies found in local rich clusters.

\smallskip (2) We have found a subset of our submm-detected QSOs 
with lower optical luminosities which
have gas and BH masses and space densities which are consistent with them being potential
`transition objects' in the proposed evolutionary scenario
where SMGs evolve into QSOs. This subset of submm-detected QSOs have
$M_\mathrm{BH}\simeq10^{8}\,M_\odot$ and gas consumption timescales
consistent with a phase of prodigious star formation where the BH has
grown substantially during the preceding $\sim200\,\mathrm{Myr}$ SMG
phase. The fast build up of the stellar population is consistent with
an evolutionary path ending in a population of present-day massive
spheroids, where homogeneous old stellar populations are seen
\citep{Swinbank06}.  

\section{Conclusions and future work}\label{concl}

We have carried out a millimetre interferometry survey of nine submm-detected QSOs at
$z=1.7$--2.6 in order to test the link between SMGs and QSOs at the era
where these two important populations were most numerous.  We include
in our analysis comparable observations of a similarly selected QSO from
the literature to provide a final sample of ten submm-detected QSOs.
To support this survey we obtained near-infrared spectroscopy
of these QSOs to derive accurate systemic redshifts needed to tune the 
millimetre receivers to the
correct frequencies.  The near-infrared spectra also provide
H$\alpha$ fluxes and line widths needed to derive reliable BH mass estimates
for the QSOs.  Our main findings are:

\smallskip (1) We detect CO emission from six of the ten submm-detected
QSOs in our sample, confirming that
they contain a significant amount of molecular gas and that a large
fraction of the mm emission is from starbursts.  The median gas mass of our sample (including
non-detections) is $(2.5\pm 0.7)\times 10^{10}$\,M$_\odot$, similar to
that found for $z\sim 2$ SMGs and to $z\gsim4$ QSOs.  
The star formation efficiencies of
our QSOs are also comparable to those measured for SMGs,
 $250\pm100$\,L$_\odot\,\mathrm{(K\,km\,s^{-1}\,pc^{2})^{-1}}$, suggesting
that the gross properties of the star formation in the QSOs are like
those seen in SMGs.   Adopting a 2\,kpc scale size for the gas
distribution in the QSOs and a typical inclination of 20$^\circ$
we derive a median dynamical mass of
M$(<2$\,kpc$)\sim(2.1\pm1.4)\times 10^{11}$\,M$_\odot$, similar
to SMGs (assuming an inclination angle appropriate for random inclinations).  
We find a lower incidence of double-peaked CO line profiles in the QSOs, compared
to SMGs, which we believe results from a selection bias towards lower
average inclination angles for the QSOs.  

\smallskip (2) Our near-infrared spectroscopy indicates a median black
hole mass in our QSO sample of $(1.8\pm1.3)\times10^{9}$\,M$_\odot$.
Combined with our dynamical estimates of the spheroid mass, these yield
$M_\mathrm{BH}/M_\mathrm{sph}\sim 9\times10^{-3}$.  This
$M_\mathrm{BH}/M_\mathrm{sph}$ ratio for this sample of submm-detected
QSOs at $z=2$ is an order of magnitude larger than the local ratio,
although $M_\mathrm{sph}$ suffers from large uncertainties 
due to the unknown CO radii and inclination angles.  This ratio is also significantly above that seen
for SMGs at $z\sim 2$.  However, this comparison masks a broad range in
BH masses within our QSO sample and so we split the sample into two
subsets based on their BH masses.

\smallskip (3) Looking at the optically luminous submm-detected QSOs in
our sample we find that we detect CO emission in 5/6 of these QSOs.
However, the estimated BH masses for these QSOs, 
$M_\mathrm{BH}\simeq10^{9}$--10$^{10}$\,M$_{\odot}$, are too large (and
their number densities too small) for them to be related to typical
SMGs in a simple evolutionary cycle.  We propose that the progenitors
of these most massive QSOs are a rare subset of SMGs with
$M_\mathrm{gas}>4\times10^{11}$\,M$_\odot$ with a number density of
$\simeq10\,\mathrm{deg}^{-2}$ which will be possible to detect with
future SCUBA-2 surveys.

\smallskip (4) For the optically less luminous ($\sim L^{\star}$)
submm-detected QSOs, we marginally detect one source in CO and obtain sensitive
limits for three further QSOs.  The BH masses for these systems are
$M_\mathrm{BH}\simeq10^{8}$\,M$_\odot$, similar to the estimates for BHs
in SMGs.  These submm-detected QSOs are consistent with being `transition' objects
between SMGs and submm-undetected QSOs, as we show it is feasible to
link their BH masses to those of SMGs by Eddington limited growth
for a period comparable to the gas depletion timescale of the QSOs,
$\sim 10$\,Myrs.  The space density of these QSOs is also in rough
agreement with that expected for the descendents of SMGs given current
estimates of the relative lifetimes of QSOs and SMGs.  We conclude that
these $\sim L^{\star}$, $M_\mathrm{BH}>10^{8}$\,M$_\odot$ submm-detected QSOs are consistent with 
being in a very brief prodigious star
formation phase, and that they simply do not possess sufficiently large gas
reservoirs to sustain the SFR (which is why these might be less often
detected in CO), although a larger sample of CO observations of
submm-detected QSOs with these BH masses is required for confirmation.

\bigskip To make further progress on understanding the evolutionary
links between SMGs and QSOs requires a larger survey of the submm and
CO emission from typical QSOs ($M_\mathrm{B}\approx -25$).  In addition, measurements of other CO
transitions for the submm-detected QSOs (e.g.~from IRAM 30-m, ALMA, 
EVLA, and SKA) are required to place better constraints on the temperature and density of the
molecular gas and thus provide a more accurate determination of the
line luminosity ratios and hence total gas masses of these
systems. Similarly, higher resolution CO observations are essential to
put strong constraints on the reservoir sizes and inclination angles, and hence $M_\mathrm{dyn}$, 
needed to compare the two populations.  Finally, better measurements of
the far-infrared SEDs (with SABOCA, SCUBA-2 or \textit{Herschel}) will
yield more accurate measures of $L_\mathrm{FIR}$ and $T_\mathrm{dust}$
for submm-detected QSOs to constrain the contribution from an AGN component.

\section{Acknowledgments}
This work is based on observations carried out with the IRAM PdBI.
IRAM is supported by INSU/CNRS (France), MPG (Germany) and IGN (Spain).
This work also makes use of data obtained at UKIRT as part of programme ID U/06A/46.  
UKIRT is operated by the Joint Astronomy Centre on behalf on the 
UK Science and Technology Facilities Council (STFC).  Ned Wright's Javascript Cosmology Calculator was 
used in preparing this paper \citep{Wright06}.

KEKC and AMS acknowledge support from the STFC.  DMA and IRS acknowledge support from the Royal Society.
We thank Thomas Greve for sending us the SMG CO spectra so that we could obtain the 
parameters of the double-Gaussian profile fits.  We thank Scott Chapman and Tadafumi Takata for allowing 
us to use their LRIS/UV and OHS/near-IR spectra of SMM\,J123716.01, SMM\,J131222.35 and MM\,J163655.
We thank Laura Hainline, David Frayer, and an anonymous referee for useful feedback which helped improve 
the clarity and presentation of the paper.

\setlength{\bibhang}{2.0em}

\end{document}